\documentclass[journal,draftclsnofoot,onecolumn,12pt,twoside]{IEEEtran}
\usepackage{algorithm} 
\usepackage{algpseudocode}
\usepackage[neverdecrease]{paralist}
\usepackage{graphicx}
\usepackage{amssymb}
\usepackage{amsfonts}
\usepackage[cmex10]{amsmath}
\usepackage{dsfont}
\usepackage{float}
\usepackage{url}
\usepackage{bm}
\usepackage{hhline}
\usepackage{enumerate}
\usepackage{pgf,tikz}
\usepackage{mathrsfs}
\usetikzlibrary{arrows}
\usetikzlibrary[patterns]

 \usepackage[colorlinks=true]{hyperref}
 \hypersetup{
    bookmarks=true,         
    unicode=false,          
    pdftoolbar=true,        
    pdfmenubar=true,        
    pdffitwindow=false,     
    pdfstartview={FitH},    
    pdftitle={My title},    
    pdfauthor={Author},     
    pdfsubject={Subject},   
    pdfcreator={Creator},   
    pdfproducer={Producer}, 
    pdfkeywords={keyword1, key2, key3}, 
    pdfnewwindow=true,      
    linkbordercolor=green,
    anchorcolor=black,filecolor=black,menucolor=black,runcolor=black,
    colorlinks=true,       
    linkcolor=red,          
    citecolor=blue,        
    filecolor=magenta,      
    urlcolor=cyan           
}

\IEEEoverridecommandlockouts
\newtheorem{Theorem}{Theorem}
\newtheorem{Proposition}{Proposition}

\newtheorem{Definition}{Definition}
\newtheorem{Example}{Example}
\newtheorem{theorem}{Theorem}

\newtheorem{lemma}[theorem]{Lemma}

\ifCLASSINFOpdf
\else
\fi
\begin{document}
\title{Construction of Full-Diversity LDPC Lattices for Block-Fading Channels}
\author{Hassan Khodaiemehr, Mohammad-Reza~Sadeghi and Daniel Panario,~\IEEEmembership{Senior Member,~IEEE}
\thanks{Hassan Khodaiemehr  and Mohammad-Reza~Sadeghi are with the Department
of Mathematics and Computer Science, Amirkabir University of Technology (Tehran Polytechnic), Tehran, Iran. Emails: \{h.khodaiemehr, msadeghi\}@aut.ac.ir.

Daniel Panario, is with the School of Mathematics and Statistics, Carleton University, Ottawa, Canada. Email: daniel@math.carleton.ca.

Part of this work has been presented in \cite{myISIT} at ISIT 2016, Spain.
}}


\maketitle
\begin{abstract}
LDPC lattices were the first family of lattices which have an efficient decoding algorithm in high dimensions over an AWGN channel.  Considering Construction D' of lattices with one binary LDPC code as underlying code gives the well known Construction A LDPC lattices or $1$-level LDPC lattices. Block-fading channel (BF) is a useful model for various
wireless communication channels in both indoor and outdoor
environments. Frequency-hopping schemes and orthogonal
frequency division multiplexing (OFDM) can conveniently
be modelled as block-fading channels. Applying lattices in this type of channel entails dividing a lattice point into
multiple blocks such that fading is constant within a block
but changes, independently,  across blocks. The design of lattices for BF channels offers a
challenging problem, which differs greatly from its counterparts like
AWGN channels. Recently, the original binary Construction A for lattices, due to Forney, have been generalized to a lattice construction from totally real and complex multiplication fields. This generalized Construction A of lattices provides signal space diversity intrinsically, which is the main requirement for the signal sets designed for fading channels.
In this paper we construct full diversity  LDPC lattices for block-fading channels using Construction A over totally real number fields. We propose a new iterative decoding method for these family of lattices which has complexity that grows linearly in the dimension of the lattice.
In order to implement our decoding algorithm, we propose the definition of a parity check matrix and Tanner graph for full diversity Construction A lattices. We also prove that the constructed LDPC lattices together with the proposed decoding method admit diversity order $n-1$ over an $n$-block-fading channel.
\end{abstract}
\section{Introduction}
\IEEEPARstart{A}{} lattice in $\mathbb {R}^{N}$ is a subgroup of $\mathbb {R}^{N}$ which is isomorphic to $\mathbb {Z}^{N}$ and spans the real vector space $\mathbb {R}^{N}$ \cite{2}. Lattices have been extensively addressed for the problem of coding in  Additive White Gaussian Noise (AWGN) channels. Communication on an AWGN channel using lattices is a communication without power constraints that has been investigated by Poltyrev~\cite{polytrev}. In such a communication system, instead of coding rate and capacity, normalized logarithmic density (NLD) and generalized capacity $C_{\infty}$ are used, respectively.

There exist different methods to construct lattices. One of the most distinguished ones is constructing lattices based on codes, where Construction A, D and D' have been proposed (for details see e.g. \cite{2}). In \cite{forneyspherebound}, it is shown that the sphere bound can be approached by a large class of coset codes or multilevel coset codes with multistage decoding,
including Construction D lattices and other certain binary lattices. Their results are based on  channel coding theorems of information theory.
As a result of their study, the concept of volume-to-noise (VNR) ratio was introduced as a parameter for measuring the efficiency of lattices  \cite{forneyspherebound}.
The subsequent challenge in lattice theory has been to find structured classes of lattices that can be encoded
and decoded with reasonable complexity in practice, and with performance that can approach the sphere-bound. This results in the transmission with arbitrary small error probability whenever VNR approaches to $1$. A capacity-achieving lattice can raise to a capacity-achieving lattice code by selecting a proper shaping region~\cite{erez,urbanke}.

Applying maximum-likelihood (ML) decoding for lattices in high dimensions  is infeasible and forced researchers to apply other low complexity decoding methods for lattices to obtain practical capacity-achieving lattices.
Integer lattices built by Construction A, D and D' can be decoded with linear complexity based on soft-decision decoding of their underlying linear binary and non-binary codes \cite{sloane,sadeghi,19,20,21,sakzad10,IWCIT2015,safarnejad}. 
The search for sphere-bound-achieving and capacity-achieving lattices and lattice codes followed by proposing low density parity-check (LDPC) lattices \cite{sadeghi},  low density lattice codes (LDLC)~\cite{LDLC} and integer low-density lattices based on Construction A (LDA)~\cite{19}. Turbo lattices, based on Construction D \cite{sakzad10}, and polar lattices \cite{polar} are other families of lattices with practical decoding methods.

Among the above family of lattices, LDPC lattices are those that have sparse parity check matrices, obtained by using a set of nested binary LDPC codes as underlying codes, together with Construction D'. If the number of underlying LDPC codes (or the level of construction) is one, Construction D'  coincides with Construction A  and $1$-level LDPC lattices are obtained \cite{IWCIT}. The theory behind Construction A is well understood.
There is a series of dualities between theoretical properties of the underlying codes and their resulting lattices. For
example there are connections between the dual of the code and the dual of the lattice, or between the weight enumerator of the code and the theta series of the lattice \cite{2,ebeling}. Construction A has been generalized
in different directions; for example a generalized construction
from the cyclotomic field $\mathbb{Q}(\xi_p)$, $\xi_p=e^{2\pi i/p}$ and $p$ a prime, is presented
in \cite{ebeling}. Then in \cite{ConstA}, a generalized construction of
lattices over a number field from linear codes is proposed.  There is consequently a rich
literature studying Construction A over different alphabets and for different tasks.

%


Lattices have  been also considered for transmission over fading channels. Specifically, algebraic lattices, defined as
lattices obtained via the ring of integers of a number field, provide efficient modulation schemes~\cite{alglattice1} for fast Rayleigh fading channels. Families of algebraic lattices are known
to reach full diversity, the first design criterion for fading
channels; see the definition of full diversity in Section \ref{system_model_subsec_2}. Algebraic lattice codes are then natural candidates for the design of codes for block-fading channels.

The block-fading channel (BF) \cite{blockfading} is a useful channel model for a class of slowly-varying wireless communication channels. Frequency-hopping schemes and orthogonal frequency division multiplexing
(OFDM), applied in many wireless communication systems standards, can conveniently
be modelled as block-fading channels.  In a BF channel a codeword spans a finite number
$n$ of independent fading blocks. As the channel realizations are constant within blocks, no codeword
is able to experience all the states of the channel; this implies that the channel is non-ergodic and therefore it is
not information stable. It follows that the Shannon capacity of this channel is zero \cite{rootLDPC}.
As far as we are aware, all available lattice based schemes on block-fading channels were proposed by using optimal decoders \cite{outage} which have exponential complexity in the worst-case. In this paper we propose full diversity LDPC lattices and their decoding method which is a mix of optimal decoding in small dimensions and iterative decoding. The proposed decoding algorithm makes it tractable to decode high-dimension LDPC lattices on the BF channel.

The rest of this paper is organized as follows. In Section~\ref{Preliminaries}, we provide preliminaries about lattices and algebraic number theory.
In Section~\ref{lattices_and_codes}, we present the available methods for constructing full diversity lattices from totally
real number fields.
The introduction of the full-diversity $1$-level LDPC lattices is also given in this section.
In Section~\ref{monogenic_sec}, the introduction of  monogenic number fields, as the tools for constructing  full-diversity $1$-level LDPC lattices, is provided.
In Section~\ref{system_model}, the system model is described for the Rayleigh block-fading channel. The available methods for evaluating the performance of finite and infinite lattice constellations over fading and block-fading channels are also discussed in this section.
In Section~\ref{new_construction}, our construction of full diversity lattices is given.
In Section~\ref{decod2}, a new decoding method is proposed for full diversity $1$-level LDPC lattices in high dimensions. The analysis of the proposed decoding method is also given in this section.
In Section~\ref{Numerical_Results}, we give computer simulations, providing decoding performance and a comparison against available bounds.
Section~\ref{conclusion} contains  concluding remarks.

\textbf{Notation}: Matrices and vectors are denoted by bold upper
and lower case letters. The $i$th element of vector $\mathbf{a}$ is denoted
by $a_i$ and the  entry $(i,j)$ of a matrix $\mathbf{A}$ is denoted by
$A_{i,j}$; $[\,\,]^t$ denotes the transposition for vectors and matrices.
\section{Preliminaries on Lattices and Algebraic Number Theory}~\label{Preliminaries}
In order to make this work self-contained, general notations and basic definitions of algebraic number theory and lattices are given next. We reveal the connection between lattices and algebraic number theory at the end of this section.
\subsection{Algebraic number theory}
Let $K$ and $L$ be two fields. If $K\subset L$, then $L$ is a field extension of $K$ denoted by $L/K$. The dimension of $L$ as vector space over $K$ is  the degree of $L$ over $K$, denoted by
$[L : K]$. Any finite extension of $\mathbb{Q}$ is  a number field.

Let $L/K$ be a field extension, and let $\alpha\in L$. If there
exists a non-zero irreducible monic  polynomial $p_{\alpha} \in K[x]$ such that $p_{\alpha}(\alpha) = 0$,  $\alpha$ is algebraic over $K$. Such a polynomial is  the minimal polynomial of $\alpha$ over $K$. If all the elements of $L$ are algebraic over $K$,  $L$ is an algebraic extension of $K$.
\begin{Definition}
Let $K$ be an algebraic number field of degree $n$; $\alpha\in K$ is an algebraic integer if it is a root of a monic polynomial with coefficients in $\mathbb{Z}$. The set of algebraic integers of $K$ is  the ring of integers of $K$, denoted by $O_K$. The ring $O_K$ is also called the maximal order of $K$.
\end{Definition}

If $K$ is a number field, then $K = \mathbb{Q}(\theta)$ for an algebraic integer $\theta\in O_K$ \cite{Stewart}. For a number field $K$ of degree $n$, the ring of integers $O_K$ forms a free $\mathbb{Z}$-module of rank $n$.
\begin{Definition}
Let $\left\{\omega_1,\ldots,\omega_n\right\}$ be a basis of the $\mathbb{Z}$-module $O_K$, so that
we can uniquely write any element of $O_K$ as $\sum_{i=1}^n a_i\omega_i$ with $a_i \in \mathbb{Z}$ for all $i$. Then, $\left\{\omega_1,\ldots,\omega_n\right\}$ is an integral basis of $K$.
\end{Definition}
\begin{Theorem}{\cite[ p. 41]{Stewart}}
Let $K = \mathbb{Q}(\theta)$ be a number field of degree $n$ over $\mathbb{Q}$. There are exactly $n$ embeddings $\sigma_1,\ldots, \sigma_n$ of $K$ into $\mathbb{C}$ defined by $\sigma_i(\theta) = \theta_i$, for $i = 1,\ldots, n$, where the $\theta_i$'s are the distinct zeros in $\mathbb{C}$ of the minimal polynomial of $\theta$ over $\mathbb{Q}$.
\end{Theorem}
\begin{Definition}
Let $K$ be a number field of degree $n$ and $x\in K$. The elements $\sigma_1(x),\ldots, \sigma_n(x)$ are
the conjugates of $x$ and
\begin{equation}\label{norm}
  N_{K/\mathbb{Q}}(x)=\prod_{i=1}^n \sigma_i(x),\quad \mathrm{Tr}_{K/\mathbb{Q}}(x)=\sum_{i=1}^n \sigma_i(x),
\end{equation}
are the norm and the trace of $x$, respectively.
\end{Definition}

For any $x \in K$, we have $N_{K/\mathbb{Q}}(x),\mathrm{Tr}_{K/\mathbb{Q}}(x)\in\mathbb{Q}$. If $x\in O_K$, we have $N_{K/\mathbb{Q}}(x),\mathrm{Tr}_{K/\mathbb{Q}}(x)\in\mathbb{Z}$.
\begin{Definition}
Let $\left\{\omega_1,\ldots,\omega_n\right\}$ be an integral basis of $K$. The discriminant of $K$ is defined as
\begin{equation}\label{disc}
d_K =\det(A)^2,
\end{equation}
where $A$ is the matrix $A_{i,j}=\sigma_j(\omega_i)$, for $i,j=1,\ldots,n$. The discriminant of a number field belongs to $\mathbb{Z}$ and it is independent of the choice of
a basis.
\end{Definition}
\begin{Definition}
Let $\left\{\sigma_1,\ldots, \sigma_n\right\}$ be the $n$ embeddings of $K$ into $\mathbb{C}$. Let $r_1$ be the number of embeddings with image in $\mathbb{R}$, the field of
real numbers, and $2r_2$ the number of embeddings with image in $\mathbb{C}$ so
that $r_1 + 2r_2 = n$. The pair $(r_1, r_2)$ is  the signature of $K$. If $r_2 = 0$ we have a totally
real algebraic number field. If $r_1 = 0$ we have a totally complex algebraic number field.
\end{Definition}
\begin{Definition}
Let us order the $\sigma_i$'s so that, for all $x\in K$, $\sigma_i(x)\in\mathbb{R}$, $1 \leq i \leq r_1$, and $\sigma_{j+r_2}(x)$ is the complex conjugate of $\sigma_j(x)$ for
$r_1 + 1 \leq j \leq r_1 + r_2$. The canonical embedding $\sigma : K \rightarrow \mathbb{R}^{r_1} \times \mathbb{C}^{r_2}$ is the homomorphism defined by
\begin{equation}\label{embeding}
  \sigma(x)=(\sigma_1(x),\ldots,\sigma_{r_1}(x),\sigma_{r_1+1}(x),\ldots, \sigma_{r_1+r_2}(x)).
\end{equation}
If we identify $\mathbb{R}^{r_1}\times \mathbb{C}^{r_2}$ with $\mathbb{R}^n$, the canonical embedding can be
rewritten as $\sigma :K \rightarrow\mathbb{R}^n$
\begin{eqnarray}\label{embeding2}
  \sigma(x)&=&(\sigma_1(x),\ldots,\sigma_{r_1}(x),\Re\sigma_{r_1+1}(x),\Im\sigma_{r_1+1}(x),\nonumber\\
  && \ldots,\Re\sigma_{r_1+r_2}(x),\Im\sigma_{r_1+r_2}(x)),
\end{eqnarray}
where $\Re$ denotes the real part and $\Im$ the imaginary part.
\end{Definition}
\begin{Definition}
A ring $A$ is  integrally closed in a field $L$ if every element
of $L$ which is integral over $A$ in fact lies in $A$. It is
integrally closed if it is integrally closed in its quotient field.
\end{Definition}
\begin{Theorem}{\cite[p. 18]{serglang}}\label{dedekind}
Let $D$ be a Noetherian ring, that is,  it satisfies the ascending chain condition on ideals, integrally closed, and such that every non-zero prime ideal is maximal. Then every ideal of $D$ can be uniquely factored into prime ideals.
\end{Theorem}

A ring satisfying the properties of Theorem~\ref{dedekind} is  a Dedekind ring.
The ring of algebraic integers in a number field is a Dedekind ring.
\begin{Definition}
Let $A$ be a ring and $x$ an element of some field $L$ containing $A$. Then, $x$ is integral over $A$ if either one of the following conditions is satisfied:
\begin{itemize}
  \item there exists a finitely generated non-zero $A$-module $M\subset L$ such
that $xM \subset M$;
  \item the element $x$ satisfies an equation
  \begin{equation*}
    x^n+a_{n-1}x^{n-1}+\cdots +a_0=0,
  \end{equation*}
with coefficients $a_i\in A$, and an integer $n\geq 1$. Such an equation
is an integral equation.
\end{itemize}
\end{Definition}
Let $A$ be a Dedekind ring, $K$ its quotient field, $L$ a finite separable
extension of $K$, and $B$ the integral closure of $A$ in $L$. If $\mathfrak{p}$ is a prime ideal
of $A$, then $pB$ is an ideal of $B$ and has a factorization
\begin{equation}\label{factorization}
\mathfrak{p}B=\mathfrak{P}_1^{e_1}\cdots \mathfrak{P}_r^{e_r},
\end{equation}
into primes of $B$, where $e_i\geq 1$. It is clear that a prime $\mathfrak{P}$ of $B$ occurs in this factorization if and only if $\mathfrak{P}$ lies above $\mathfrak{p}$. Each $e_i$ is called the ramification index of $\mathfrak{P}_i$ over $\mathfrak{p}$, and is also written $e(\mathfrak{P}_i/\mathfrak{p})$.
If $\mathfrak{P}$ lies above $\mathfrak{p}$ in $B$, we denote by $f(\mathfrak{P}/\mathfrak{p})$ the degree of the residue class field extension $B/\mathfrak{P}$ over $A/\mathfrak{p}$, and call it the residue class degree or inertia degree.
\begin{Theorem}{\cite[p. 24]{serglang}}
Let $A$ be a Dedekind ring, $K$ its quotient field, $L$ a
finite separable extension of $K$, and $B$ the integral closure of $A$ in $L$. Let
$\mathfrak{p}$ be a prime of $A$. Then
\begin{equation}\label{ramification}
[L:K]=\sum_{\mathfrak{P}|\mathfrak{p}}e(\mathfrak{P}/\mathfrak{p}) f(\mathfrak{P}/\mathfrak{p}).
\end{equation}
\end{Theorem}
When $L/K$ is a Galois extension of degree $n$, this simplifies to $n =efg$, where $g$ is the number of primes $\mathfrak{P}$ of $B$ above $\mathfrak{p}$. In other words, $e(\mathfrak{P}/\mathfrak{p}) = e$ and $f(\mathfrak{P}/\mathfrak{p}) = f$ for all $\mathfrak{P}|\mathfrak{p}$. If $e_{\mathfrak{P}} =f_{\mathfrak{P}}$  for all $\mathfrak{P}|\mathfrak{p}$, then  $\mathfrak{p}$ \emph{splits completely} in $L$. In that case, there are exactly $[L: K]$ primes of $B$ lying above $\mathfrak{p}$. A prime $\mathfrak{p}$ in $K$ is \emph{ramified} in a number field $L$  if the prime ideal factorization~(\ref{factorization}) has some $e_i$ greater than $1$. If every $e_i$ equals $1$, $\mathfrak{p}$ is \emph{unramified} in $L$.
If $[L: K] = e(\mathfrak{P}/\mathfrak{p})$, $\mathfrak{P}$ is \emph{totally ramified} above $\mathfrak{p}$. In this case, the residue class degree is equal to $1$. Since $\mathfrak{P}$ is the only prime of $B$ lying above $\mathfrak{p}$,  $L$ is \emph{totally ramified} over $K$.  If the characteristic $p$ of the residue
class field $A/\mathfrak{p}$ does not divide $e(\mathfrak{P}/\mathfrak{p})$, then $\mathfrak{P}$ is \emph{tamely ramified} over $\mathfrak{p}$ (or $L$ is tamely ramified over $K$). If it does, then $\mathfrak{P}$ is \emph{strongly ramified}.
\subsection{Lattices}~\label{A.Lattices}
Any discrete additive subgroup $\Lambda$ of the $m$-dimensional real space $\mathbb{R}^m$ is a lattice.  Every lattice $\Lambda$ has a basis $\mathcal{B}=\{{\mathbf b}_1,\ldots,{\mathbf b}_n\}\subseteq\mathbb{R}^m$,  $n \leq m$, where every ${\mathbf x}\in\Lambda$ can be represented as an integer linear combination of vectors in $\mathcal{B}$. The $n\times m$ matrix $\mathbf{M}$ with ${\mathbf b}_1,\ldots,{\mathbf b}_n$ as rows, is a generator matrix for the lattice.
The rank of the lattice is $n$ and its dimension is $m$.
If $n = m$, the lattice is  a full-rank lattice. In this paper, we consider only full-rank lattices.
A lattice $\Lambda$  can be described in terms of a generator matrix $\mathbf{M}$ by
\begin{equation}\label{lattice_def}
  \Lambda=\left\{\mathbf{x}=\mathbf{uM}\,|\,\mathbf{u}\in\mathbb{Z}^{n}\right\}.
\end{equation}
When using lattices for coding, their Voronoi cells and volume always play an important role. For any lattice
point $\mathbf{p}$ of a lattice $\Lambda\subset \mathbb{R}^m$, its Voronoi cell is defined by
\begin{equation}
\mathcal{V}_{\Lambda}(\mathbf{p})=\left\{\mathbf{x}\in \mathbb{R}^{m},\,d(\mathbf{x},\mathbf{p})\leq d(\mathbf{x},\mathbf{q})\,\, \textrm{for}\,\, \textrm{all} \,\, \mathbf{q}\in\Lambda\right\}.
\end{equation}
All Voronoi cells are the same, thus $\mathcal{V}_{\Lambda}(\mathbf{p})=\mathcal{V}_{\Lambda}(\mathbf{0})\triangleq \mathcal{V}(\Lambda)$.
The matrix $\mathbf{G} = \mathbf{M}\mathbf{M}^t$ is  a Gram matrix for the lattice. The determinant of the lattice $\det(\Lambda)$ is defined as the determinant of the matrix $\mathbf{G}$ and the volume of the lattice is
\begin{eqnarray}\label{volume}
  \textrm{vol}(\Lambda)=\textrm{vol}(\mathcal{V}(\Lambda))=\sqrt{\det(\mathbf{G})}.
\end{eqnarray}
\begin{Definition}
A lattice $\Lambda$ in $\mathbb{R}^m$ is an integral lattice if its Gram matrix has coefficients in $\mathbb{Z}$. Indeed, a lattice $\Lambda$ is integral if and only if $\left\langle x, y\right\rangle\in \mathbb{Z}$, for all $x, y \in \Lambda$, where $\left\langle ,\right\rangle$ is the regular inner product of $\mathbb{R}^m$.
\end{Definition}

The set of all vectors in $\mathbb{R}^m$ whose inner product with all vectors of $\Lambda$ is in $\mathbb{Z}$ is another lattice, the  dual lattice of $\Lambda$, denoted by $\Lambda^{\ast}$.
The \emph{normalized volume} of an $n$-dimensional lattice $\Lambda$ is defined as $\det(\Lambda)^{2/n}$~\cite{forneyspherebound}. This volume may be regarded as the volume of $\Lambda$ per two dimensions.

Suppose that the points of a lattice $\Lambda$ are sent over an unconstrained Additive White Gaussian Noise (AWGN)~\cite{polytrev} channel, with noise variance $\sigma^2$. Let the vector $\mathbf{x}\in\Lambda$ be transmitted over the unconstrained AWGN channel, then the received vector $\mathbf{r}$ can be written as $\mathbf{r}=\mathbf{x}+\mathbf{e}$,
where $\mathbf{e}=(e_1,\ldots,e_n)$ is the error term and its components are
independently and identically distributed  (i.i.d.) with $\mathcal{N}(0,\sigma^2)$.
The \emph{volume-to-noise ratio} (VNR) of the lattice $\Lambda$ is defined as
\begin{equation}~\label{VNR}
{\mbox {VNR}}=\frac{\textrm{vol}(\Lambda)^{\frac{2}{n}}}{2\pi e\sigma^2}.
\end{equation}
For large $n$, VNR is the ratio of the normalized volume of $\Lambda$ to the normalized volume of a noise sphere of squared radius $n\sigma^2$ which is defined as generalized signal-to-noise ratio (SNR) in~\cite{sadeghi} and $\alpha^2$ in~\cite{forneyspherebound}. Due to the geometric uniformity of lattices, the probability of error under maximum likelihood decoding of $\Lambda$ is the probability that a white Gaussian $n$-tuple ${\mathbf r}$ with noise variance $\sigma^2$ falls outside the Voronoi cell $\mathcal{V}({\mathbf 0})=\mathcal{V}$.

Now we present definitions in algebraic lattice theory  equivalent to the above definitions.
\begin{Definition}
An integral lattice $\Gamma$ is a free $\mathbb{Z}$-module of finite rank together with a positive definite symmetric bilinear form $\left\langle ,\right\rangle:\Gamma\times \Gamma \rightarrow \mathbb{Z}$.
\end{Definition}
\begin{Definition}
The discriminant of a lattice $\Gamma$, denoted
$\mathrm{disc}(\Gamma)$, is the determinant of $\mathbf{M}\mathbf{M}^t$ where $\mathbf{M}$ is a generator matrix for $\Gamma$. The volume $\textrm{vol}(\mathbb{R}^n/\Gamma)$ of a lattice $\Gamma$ is defined as $|\det(\mathbf{M})|$.
\end{Definition}

The discriminant is related to the volume of a lattice by
\begin{equation}\label{disc}
  \textrm{vol}(\mathbb{R}^n/\Gamma)=\sqrt{\mathrm{disc}(\Gamma)}.
\end{equation}
Moreover, when $\Gamma$ is integral, we have $\mathrm{disc}(\Gamma) = |\Gamma^{*}/\Gamma|$, where $\Gamma^{*}$ is the dual of the lattice $\Gamma$ defined by
\begin{equation}\label{dual}
  \Gamma^{*}=\left\{y\in \mathbb{R}^m\,\,|\,\,y\cdot x\in \mathbb{Z} \,\,\,\mbox{for}\,\,\mbox{all}\,\,\, x\in \Gamma\right\}.
\end{equation}
When $\Gamma = \Gamma^{*}$, the lattice $\Gamma$ is unimodular.

The canonical embedding~(\ref{embeding2})  gives a geometrical representation of a number
field and makes the connection between algebraic number fields and lattices.
\begin{Theorem}{\cite[p. 155]{Stewart}}
Let $\left\{\omega_1, \omega_2,\ldots,\omega_n\right\}$ be an integral basis of a number field $K$. The $n$ vectors $\mathbf{v}_i = \sigma(\omega_i) \in \mathbb{R}^n$, $i = 1,\ldots,n$ are linearly independent,
so they define a full rank lattice $\Lambda = \Lambda(O_K) = \sigma(O_K)$.
\end{Theorem}
\begin{Theorem}{\cite{samuel}}\label{vol_theorem}
Let $d_K$ be the discriminant of a number field $K$. The volume of the fundamental parallelotope of $\Lambda(O_K)$ is given by
\begin{equation}\label{vol_alg}
  \textrm{vol}(\Lambda(O_K))=2^{-r_2}\sqrt{|d_K|}.
\end{equation}
\end{Theorem}
\section{Lattice Constructions using Codes} \label{lattices_and_codes}
There exist many ways to construct lattices based on codes~\cite{2}. Here we mention  a lattice construction from totally real and complex multiplication fields \cite{ConstA}, which naturally generalizes
Construction A of lattices from $p$-ary codes obtained from the
cyclotomic field $\mathbb{Q}(\xi_p)$, with $\xi_p=e^{2\pi i/p}$ and $p$ a prime number \cite{ebeling}. This contains the
so-called Construction A of lattices from binary codes as a particular case.
\subsection{Construction A of lattices}~\label{algebraic_const_A}
Given a number field $K$ and a prime $\mathfrak{p}$  of $\mathcal{O}_K$ above $p$ where $\mathcal{O}_K/\mathfrak{p}\cong \mathbb{F}_{p^f}$, let $\mathcal{C}$ be an $(N, k)$ linear code over $\mathbb{F}_{p^f}$. The Construction A of lattices using underlying code $\mathcal{C}$ and number field $K$ is given in \cite{ConstA}.
\begin{Definition}
Let $\rho : \mathcal{O}_K^N\rightarrow \mathbb{F}_{p^f}^{N}$ be the mapping defined by the reduction modulo the ideal $\mathfrak{p}$ in each of the $N$ coordinates. Define $\Gamma_{\mathcal{C}}$ to be the preimage of $\mathcal{C}$ in $\mathcal{O}_K^N$, i.e.,
\begin{equation}\label{alg_costA}
\Gamma_{\mathcal{C}}=\left\{\mathbf{x}\in \mathcal{O}_K^N\,\,|\,\, \rho(\mathbf{x})=\mathbf{c},\,\,  \mathbf{c}\in \mathcal{C}\right\}.
\end{equation}
\end{Definition}

We conclude that $\Gamma_{\mathcal{C}}$ is a $\mathbb{Z}$-module of rank $nN$.
When $K$ is totally real, $\rho^{-1}(\mathcal{C})$ forms a lattice with the following symmetric bilinear form \cite{ConstA}
\vspace{-0.2cm}
\begin{equation}\label{bilinear}
  \left\langle x,y\right\rangle=\sum_{i=1}^N \mbox{Tr}_{K/\mathbb{Q}}(\alpha x_i y_i),
\end{equation}
where $\mathbf{x}=(x_1,\ldots,x_N)$ and $\mathbf{y}=(y_1,\ldots, y_N)$ are vectors in $\mathcal{O}_K^N$, $\alpha\in \mathcal{O}_K$ is a totally positive element, meaning that $\sigma_i(\alpha)> 0$ for all $i$, and $\mbox{Tr}_{K/\mathbb{Q}}$ is defined in (\ref{norm}). Thus, $\Gamma_{\mathbf{C}}$ together with the bilinear
form~(\ref{bilinear}) is an integral lattice. A similar construction is obtained from a CM-field \cite{ConstA}. A
CM-field is a totally imaginary quadratic extension of a totally
real number field. If $K$ is a CM-field and $\alpha \in  O_K \cap \mathbb{R}$ is totally positive, then $\rho^{-1}(\mathcal{C})$ forms a lattice with the following symmetric bilinear form
\begin{equation}\label{bilinear2}
  \left\langle x,y\right\rangle=\sum_{i=1}^N \mathrm{Tr}_{K/\mathbb{Q}}(\alpha x_i \bar{y_i}),
\end{equation}
where $\bar{y_i}$ denotes the complex conjugate of $y_i$. If $K$ is totally real, then $\bar{y_i}=y_i$, and this
notation treats both cases of totally real and CM-fields at the same time. It has been shown that if $\mathcal{C}\subset \mathcal{C}^{\bot}$, then $\sum_{i=1}^N \mathrm{Tr}_{K/\mathbb{Q}}(x_i \bar{y_i})\in p\mathbb{Z}$, and thus  the symmetric bilinear form can be normalized by a factor $1/p$, or equivalently, by choosing $\alpha = 1/p$ \cite{ConstA}.

Other variations of the above construction have been considered
in the literature. The case $N = 1$ is considered in \cite{ref18} where  the problem reduces to understanding which lattices can be obtained on the ring of integers of a number field. The case  that $K$ is the cyclotomic field $\mathbb{Q}(\xi_p)$ has been considered in \cite{ebeling}. In \cite{ref12}, the prime ideal $\mathfrak{p}$ is considered to be $(2m)$, yielding codes over a ring of polynomials
with coefficients modulo $2m$. In \cite{ref19}, $\mathfrak{p}$ is considered to be $(2-\xi_p+\xi_p^{-1})$ and the resulting codes are over $\mathbb{F}_p$. Quadratic extensions $K = \mathbb{Q}(\sqrt{-l})$ are
considered in \cite{ref13} and \cite{ref20} where the reduction is done by the ideal $(p^e)$ and the resulting codes are over the ring $O_K/p^eO_K$.

A generator matrix for the lattice $\Gamma_{\mathcal{C}}$ is computed in \cite{ConstA}. Let $K$ be a Galois extension and
the prime $\mathfrak{p}$ be chosen so that $\mathfrak{p}$ is totally ramified. Therefore, we have $pO_K = \mathfrak{p}^n$. Now, let $\mathcal{C} \subset \mathbb{F}_p^N$ be a linear code over $\mathbb{F}_p$ of length $N$.
Since $\Gamma_{\mathcal{C}}$ has rank $nN$ as a free $\mathbb{Z}$-module, we  obtain the $\mathbb{Z}$-basis of $\Gamma_{\mathcal{C}}$. Let $\left\{\omega_1,\ldots,\omega_n\right\}$ be a $\mathbb{Z}$-basis of $O_K$. Then,
a generator matrix for the lattice formed by $O_K$ together with
the standard trace form $\left\langle w, z\right\rangle = Tr_{K/\mathbb{Q}}(wz)$, $w, z \in O_K$, is given by
\begin{equation}\label{gen_OK}
  \mathbf{M}=[\sigma_j(\omega_i)]_{i,j=1}^{n}.
\end{equation}
The prime ideal $\mathfrak{p}$ is a $\mathbb{Z}$-module of rank $n$. It then has a $\mathbb{Z}$-basis $\left\{\mu_1,\ldots, \mu_n\right\}$ where $\mu_i=\sum_{j=1}^n \mu_{i,j}\omega_j$. Thus
\begin{equation}\label{gen_P}
  [\sigma_j(\mu_i)]_{i,j=1}^{n}=\mathbf{D}\mathbf{M},
\end{equation}
where $\mathbf{D}=[\mu_{i,j}]_{i,j=1}^{n}$.
\begin{Theorem}{\cite[Proposition 1]{ConstA}}\label{theorem6}
The lattice $\Gamma_{\mathcal{C}}$ is a sublattice of $O_K^N$
with discriminant
\begin{equation}\label{disc}
  \mathrm{disc}(\Gamma_{\mathcal{C}})=d_K^N(p^f)^{2(N-k)},
  \end{equation}
where $d_K = (\det([\sigma_i(\omega_j)]_{i,j=1}^n))^2$ is the discriminant of $K$. The
lattice $\Gamma_{\mathcal{C}}$ is given by the generator matrix
\begin{equation}\label{Gamma_C_gen}
  \mathbf{M}_{\mathcal{C}}=\left[
                 \begin{array}{cc}
                   \mathbf{I}_k\otimes \mathbf{M} & \mathbf{A}\otimes \mathbf{M} \\
                   \mathbf{0}_{n(N-k)\times nk} & \mathbf{I}_{N-k}\otimes \mathbf{DM} \\
                 \end{array}
               \right],
\end{equation}
where $\otimes$ is the tensor product of matrices, $\left[
                                                      \begin{array}{cc}
                                                        \mathbf{I}_k & \mathbf{A} \\
                                                      \end{array}
                                                    \right]
$ is a generator matrix of ${\mathcal{C}}$, $\mathbf{M}$ is the matrix of embeddings of a $\mathbb{Z}$-basis of $O_K$ given in~(\ref{gen_OK}), and $\mathbf{DM}$ is the matrix of embeddings of a $\mathbb{Z}$-basis of $\mathfrak{p}$ in~(\ref{gen_P}).
\end{Theorem}

\subsection{LDPC lattices from Construction A and Construction \textrm{D'}}~\label{ConstructionAversusConstructionD'}
Assume that $\mathcal{C}$ is a linear code over $\mathbb{F}_p$ where $p$ is a prime number, i.e. $\mathcal{C}\subseteq\mathbb{F}_{p}^N$. Let $d_{\min}$ denote the minimum distance of $\mathcal{C}$. A lattice $\Lambda$ constructed based on Construction A \cite{2} can be derived from $\mathcal{C}$ by:
\begin{equation}\label{constA}
\Lambda=p\mathbb{Z}^N+\epsilon\left(\mathcal{C}\right),
\end{equation}
where $\epsilon\colon\mathbb{F}_p^N\rightarrow\mathbb{R}^N$ is the embedding function which sends a vector in $\mathbb{F}_p^N$ to its real version. In this work, we are particularly interested in binary codes and lattices with $p=2$.

Construction D' converts a set of parity checks defined by a family of nested codes into congruences for a lattice~\cite{2}. This construction is a good tool to produce lattices based on LDPC codes. Let $\mathcal{C}_0\supseteq \mathcal{C}_1\supseteq \cdots \supseteq \mathcal{C}_a$ be a family of nested linear codes, where $\mathcal{C}_\ell$ has parameter $\left[n,k_\ell,d_{\min}^\ell \right]$, for $0\leq \ell\leq a$. Let $\{{\mathbf h}_1,\ldots ,{\mathbf h}_N\}$ be a linearly independent set of vectors in $\mathbb{F}_2^N$ and the code $\mathcal{C}_\ell$ be defined by the $r_\ell=N-k_\ell$ parity check vectors ${\mathbf h_1},\ldots ,{\mathbf h}_{r_\ell}$.
Define the new lattice $\Lambda$ consisting of those ${\mathbf x}\in \mathbb{Z}^N$ that satisfy the congruences ${\mathbf h}_j\cdot {\mathbf x}\equiv 0 \pmod{2^{\ell+1}}$ for $0\leq \ell\leq a$ and $r_{a-\ell-1}+1\leq j\leq r_{a-\ell}$. The number $a+1$ is the level of the construction.

By multiplying the modular equations by appropriate powers of $2$, we can restate Construction D'~\cite{sadeghi}. Indeed, ${\mathbf x}\in\Lambda$ if and only if  ${\mathbf H}_{\Lambda}{\mathbf x}^t=\mathbf{0} \pmod{2^{a+1}}$ where
\begin{equation}~\label{H}
\mathbf{H}_{\Lambda}=[{\mathbf h}_1,\ldots,{\mathbf h}_{r_0},\ldots,2^a{\mathbf h}_{r_{a-1}+1},\ldots,2^a{\mathbf h}_{r_{a}}]^t.
\end{equation}
Then, ${\mathbf H}_{\Lambda}$ constitutes the parity check matrix of $\Lambda$. When the underlying codes are binary LDPC codes, the lattice $\Lambda$ constructed based on Construction D' and associated to this ${\mathbf H}_{\Lambda}$ is  an $(a+1)$-level LDPC lattice. The Tanner graph of these lattices can be constructed based on their parity check matrices ${\mathbf H}_{\Lambda}$ and used for decoding purposes \cite{sadeghi}.
\begin{Definition}
A $1$-level LDPC lattice $\Lambda$ is a lattice based on Construction D' along with a binary linear LDPC code $\mathcal{C}$ as its underlying code. Equivalently, ${\mathbf x}\in\mathbb{Z}^N$ is in $\Lambda$ if $\mathbf H_{\mathcal{C}}{\mathbf x}^t=\mathbf{0} \pmod{2}$, where ${\mathbf H}_{\mathcal{C}}$ is the parity-check matrix of $\mathcal{C}$ \cite{IWCIT2015,IWCIT,QC_LDPC_lattice}.
\end{Definition}
\begin{Proposition}~\label{th:equivalenD'A}
A $1$-level LDPC lattice $\Lambda$ is equal to a lattice $\Lambda_1$ constructed following Construction A using the same underlying code $\mathcal{C}$ \cite{IWCIT}.
\end{Proposition}

The generator matrix of $1$-level LDPC lattice $\Lambda$ using the underlying code $\mathcal{C}$ is of the form \cite{2,IWCIT2015}:
\begin{eqnarray}\label{eq12}
  \mathbf{G}_{\Lambda} &=& \left[
                    \begin{array}{cc}
                      \mathbf{I}_{k}& \mathbf{P}_{k\times (N-k)} \\
                      \mathbf{0}_{(N-k)\times k} & 2\mathbf{I}_{N-k} \\
                    \end{array}
                  \right],
\end{eqnarray}
\noindent
where $\mathbf{G}_{\mathcal{C}}=\left[
                           \begin{array}{cc}
                              \mathbf{I}_k & \mathbf{P}_{k\times (N-k)}\\
                           \end{array}
                         \right]$
is the generator matrix of $\mathcal{C}$ in systematic form, $k$ is the rank of $\mathcal{C}$ and $N$ is the code length of $\mathcal{C}$.
The matrices $\mathbf{I}_k$ and $\mathbf{0}_{(N-k)\times k}$ are the identity matrix of size $k$ and the all zero  matrix of size $(N-k)\times k$, respectively.
\begin{Example}\label{example1}
This example is discussed in \cite{ConstA}. Let $p$ be an odd prime, and let $\xi_p$ be a primitive $p$th root of unity. Consider the cyclotomic field $K = \mathbb{Q}(\xi_p)$ with the ring of integers $O_K = \mathbb{Z}[\xi_p]$. The
degree of $K$ over $\mathbb{Q}$ is $p - 1$, and $p$ is totally ramified, with $pO_K = (1-\xi_p)^{p-1}$. Thus, taking the prime ideal $\mathfrak{p} = (1-\xi_p)$ with the residue field $O_K /\mathfrak{p}\simeq \mathbb{F}_p$,  the bilinear form $\left\langle x,y\right\rangle=\sum_{i=1}^N \mathrm{Tr}_{K/Q}(x_iy_i)$ and a linear code $\mathcal{C}$ over $\mathbb{F}_p$, then $\Gamma_{\mathcal{C}}$ yields the so-called Construction A as described above.
Since $\mathbb{Q}(\xi_p)$ is a CM-field, we can  use the  bilinear form corresponding to~(\ref{bilinear2}) with $\alpha = 1/p$. By using this bilinear form, the generator matrix is as follows
\begin{eqnarray}\label{constA_p}
  \mathbf{M}_{\mathcal{C}} &=& \frac{1}{\sqrt{p}}\left[
                    \begin{array}{cc}
                      \mathbf{I}_{k}& \mathbf{P}_{k\times (N-k)} \\
                      \mathbf{0}_{(N-k)\times k} & p\mathbf{I}_{N-k} \\
                    \end{array}
                  \right].
\end{eqnarray}
It has been proved in \cite{ConstA} that if $\mathcal{C} \subset \mathcal{C}^{\perp}$, then $\Gamma_{\mathcal{C}}$ is an integral lattice of rank $N(p - 1)$.
Our particular case is based on Construction A of lattices from codes  when $p = 2$.
In such case, $\xi_p = -1$, $O_K = \mathbb{Z}$, and $\mathfrak{p} = 2\mathbb{Z}$.
$\hfill\square$\end{Example}

Next, we present the definition of $1$-level LDPC lattices using algebraic number fields.
\begin{Definition}\label{parity_def}
Let $\mathcal{C}$ be a binary LDPC code of length $N$ and dimension $k$.  Consider the number field $K$ with the ring of integers $O_K$. Let $n$ be the degree of $K$ over $\mathbb{Q}$ and $\mathfrak{p}$ be a prime in $O_K$ with residue field $O_K /\mathfrak{p}\simeq \mathbb{F}_2$. Define $\rho:O_K^N\rightarrow \mathbb{F}_2^N $ as the componentwise reduction modulo $\mathfrak{p}$ and $\sigma^i:O_K^i\rightarrow \mathbb{R}^{in}$, for positive integer $i$, as
$$\sigma^i(x_1,\ldots,x_i)=(\sigma(x_1),\ldots,\sigma(x_i)),$$
where $\sigma$ is the canonical embedding in~(\ref{embeding2}). Let $\left\{\omega_1,\ldots,\omega_n\right\}$ be the integral basis for $O_K$. Define $\sigma^{-1}:\sigma(O_K)\rightarrow O_K$ such that for $x=\sum_{l=1}^n u_{i,l}\omega_i$ in $O_K$
$$\sigma^{-1}(\sigma_1(x),\ldots,\sigma_n(x))=x.$$
Define $(\sigma^i)^{-1}$ similarly to $\sigma^i$ but replacing $\sigma$ with $\sigma^{-1}$.
Then, $\Lambda=\sigma^N(\Gamma_{\mathcal{C}})=\sigma^N(\rho^{-1}(\mathcal{C}))$ is  the $1$-level LDPC lattice based on the number field $K$. The parity check matrix $\mathbf{H}_{\Lambda}$ for $\Lambda$ is an $n(N-k)\times nN$ matrix over $\mathbb{F}_2$ of rank $n(N-k)$ such that
\begin{equation}\label{parity_check}
  \Lambda=\left\{\mathbf{x}\in\sigma^{N}(O_K)\,\,|\,\,\rho((\sigma^{N-k})^{-1}(\mathbf{x}\mathbf{H}^t))=\mathbf{0}_{1\times (N-k)}\right\}.
\end{equation}
\end{Definition}
\begin{Theorem}\label{theorem_parity}
Let $\mathcal{C}$ be a binary LDPC code of length $N$ and dimension $k$. Let $\mathbf{H}$ and $\mathbf{G}=\left[
                           \begin{array}{cc}
                              \mathbf{I}_k & \mathbf{A}\\
                           \end{array}
                         \right]$
be the parity check and generator matrices of $\mathcal{C}$, respectively. Consider the Galois extension $K/\mathbb{Q}$ with the ring of integers $O_K$. Let $n$ be the degree of $K$ over $\mathbb{Q}$ and let $2$ be totally ramified in $O_K$. The prime $\mathfrak{p}$  is chosen above $2$ so that $2O_k=\mathfrak{p}^n$ with  residue field $O_K /\mathfrak{p}\simeq \mathbb{F}_2$. Then, $\mathbf{H}_{\Lambda}=\mathbf{H}\otimes\mathbf{I}_n$ is the parity check matrix of $1$-level LDPC lattices $\Lambda=\sigma^N(\Gamma_{\mathcal{C}})=\sigma^N(\rho^{-1}(\mathcal{C}))$.
\end{Theorem}
\begin{IEEEproof}
Based on the assumed conditions and Theorem~\ref{theorem6}, the generator matrix of $\Lambda$ has the following form
 \begin{equation*}
  \mathbf{M}_{\Lambda}=\left[
                 \begin{array}{cc}
                   \mathbf{I}_k\otimes \mathbf{M} & \mathbf{A}\otimes \mathbf{M} \\
                   \mathbf{0}_{n(N-k)\times nk} & \mathbf{I}_{N-k}\otimes \mathbf{DM} \\
                 \end{array}
               \right].
\end{equation*}
Let $\mathbf{u}=(u_1,\ldots,u_{nN})$ be an integer vector. First we show that $\rho((\sigma^{N-k})^{-1}(\mathbf{u}\mathbf{M}_{\Lambda}\mathbf{H}_{\Lambda}^t))=\mathbf{0}$. To this end,
\begin{eqnarray*}
   \mathbf{M}_{\Lambda}\mathbf{H}_{\Lambda}^t &=&
   \left[
                 \begin{array}{c}
                   \left[\mathbf{I}_k \,\,\,\,\, \mathbf{A}\right]\otimes \mathbf{M} \\
                   \left[\mathbf{0}_{(N-k)\times k} \,\,\, \mathbf{I}_{N-k}\right]\otimes \mathbf{DM}\\
                 \end{array}
   \right](\mathbf{H}\otimes\mathbf{I}_n)^t \\
   &=& \left[
                 \begin{array}{c}
                   \left[\mathbf{I}_k \,\,\,\,\, \mathbf{A}\right]\mathbf{H}^t\otimes \mathbf{M} \\
                   \left[\mathbf{0}_{(N-k)\times k} \,\,\, \mathbf{I}_{N-k}\right]\mathbf{H}^t\otimes \mathbf{DM}\\
                 \end{array}
   \right].
\end{eqnarray*}
The $\mathbb{Z}$-linearity of $(\sigma^{N-k})^{-1}$ implies the sufficiency of proving  $\rho((\sigma^{N-k})^{-1}(\mathbf{b}_i))=\mathbf{0}$, where $\mathbf{b}_i$ is the $i$th row of $\mathbf{M}_{\Lambda}\mathbf{H}_{\Lambda}^t$, for $i=1,\ldots,nN$. Since $\mathbf{H}$ and $\left[\mathbf{I}_k \,\, \mathbf{A}\right]$ are the parity check matrix and the generator matrix of the binary code $\mathcal{C}$, respectively, $\left[\mathbf{I}_k \,\, \mathbf{A}\right]\mathbf{H}^t=2\mathbf{Z}$ for a $k\times(N-k)$ integer matrix $\mathbf{Z}$. On the other hand, $\left[\mathbf{0}_{(N-k)\times k} \,\, \mathbf{I}_{N-k}\right]\mathbf{H}^t= \mathbf{H}_{N-k}$, where $\mathbf{H}_{N-k}$ is the last $N-k$ rows of $\mathbf{H}^t$. For $1\leq i\leq kn$, let $r_i=\left\lfloor\frac{i}{n}\right\rfloor+1$, where $\left\lfloor c\right\rfloor$ is the floor of a real number $c$, and $s_i=i-(r_i-1)n$. Then
$$\mathbf{b}_i=\left( 2z_{r_i,1}\mathbf{M}_{s_i} , 2z_{r_i,2}\mathbf{M}_{s_i},  \ldots , 2z_{r_i,N-k}\mathbf{M}_{s_i}
               \right),$$
in which $\mathbf{Z}_{r_i}=\left(z_{r_i,1},\ldots,z_{r_i,N-k}\right)$ and $\mathbf{M}_{s_i}=\left(\sigma_1(\omega_{s_i}),\ldots,\sigma_n(\omega_{s_i})\right)$ are $r_i$th and $s_i$th rows of $\mathbf{Z}$ and $\mathbf{M}$, respectively. Finally,
\begin{IEEEeqnarray*}{rCl}
  && \rho((\sigma^{N-k})^{-1}(\mathbf{b}_i)) \\
  &=& \rho((\sigma^{N-k})^{-1}\left(
                   2z_{r_i,1}\mathbf{M}_{s_i} ,  \ldots , 2z_{r_i,N-k}\mathbf{M}_{s_i}
               \right)) \\
   &=& \rho\left(
                   2z_{r_i,1}\sigma^{-1}(\mathbf{M}_{s_i}),  \ldots , 2z_{r_i,N-k}\sigma^{-1}(\mathbf{M}_{s_i})
               \right)  \\
   &=& \rho\left(
                   2z_{r_i,1}\omega_{s_i} ,  \ldots , 2z_{r_i,N-k}\omega_{s_i}
               \right) \\
   &=&\mathbf{0},
\end{IEEEeqnarray*}
where the last equation follows from the fact that
$$\left(2z_{r_i,1}\omega_{s_i},  \ldots,  2z_{r_i,N-k}\omega_{s_i}
               \right) \in (2O_K)^{N-k}\subset \mathfrak{p}^{N-k}.$$
For $kn+1\leq i\leq nN$,  let $r_i=\left\lfloor\frac{i}{n}\right\rfloor-k+1$,  and $s_i=i-(r_i+k-1)n$. Consider $\left\{\mu_1,\ldots,\mu_n\right\}$ as the $\mathbb{Z}$-basis of $\mathfrak{p}$.  Then
$$\mathbf{b}_i=\left( h_{r_i,1}\mathbf{P}_{s_i} , h_{r_i,2}\mathbf{P}_{s_i} , \ldots , h_{r_i,N-k}\mathbf{P}_{s_i}
               \right),$$
where $(h_{r_i,1},\ldots,h_{r_i,N-k})$ and $\mathbf{P}_{s_i}=(\sigma_1(\mu_{s_i}),\ldots,\sigma_n(\mu_{s_i}))$ are the $r_i$th and $s_i$th rows of $\mathbf{H}_{N-k}$ and $\mathbf{DM}$, respectively. In this case
\begin{IEEEeqnarray*}{rCl}
  &&\rho((\sigma^{N-k})^{-1}(\mathbf{b}_i))\\&=& \rho((\sigma^{N-k})^{-1}\left(
                   h_{r_i,1}\mathbf{P}_{s_i} ,  \ldots , h_{r_i,N-k}\mathbf{P}_{s_i}
               \right)) \\
   &=& \rho\left(
                   h_{r_i,1}\sigma^{-1}(\mathbf{P}_{s_i}),  \ldots , h_{r_i,N-k}\sigma^{-1}(\mathbf{P}_{s_i})
               \right)  \\
   &=& \rho\left(
                   h_{r_i,1}\mu_{s_i} ,  \ldots , h_{r_i,N-k}\mu_{s_i}
               \right) \\
   &=&\mathbf{0}.
\end{IEEEeqnarray*}
Now, let $\mathbf{x}\in\sigma^{N}(O_K)$ such that $\rho((\sigma^{N-k})^{-1}(\mathbf{x}\mathbf{H}_{\Lambda}^t))=\mathbf{0}$. We  show that $\mathbf{x}\in\Lambda$. For the sake of this, we have
$$\mathbf{x}=\left(\sigma_1(x_{1}),\ldots,\sigma_n(x_{1}),\ldots,\sigma_1(x_{N}),\ldots,\sigma_n(x_{N})\right),$$
where $\tilde{\mathbf{x}}=(x_1,\ldots,x_N)\in O_K^N$. Then
\begin{IEEEeqnarray*}{rCl}
  \mathbf{x}\mathbf{H}_{\Lambda}^t &=& \mathbf{x}\left[\mathbf{h}_1 , \mathbf{h}_2 , \ldots , \mathbf{h}_{n(N-k)}
                                                                       \right]^t
   \\
   &=& \left(\mathbf{x}\cdot \mathbf{h}_1^t,\mathbf{x}\cdot \mathbf{h}_1^t,\ldots,\mathbf{x}\cdot \mathbf{h}_{n(N-k)}^t\right),
\end{IEEEeqnarray*}
where $\mathbf{x}\cdot \mathbf{h}_i^t$ is the inner product of $\mathbf{x}$ and the $i$th column of $\mathbf{H}_{\Lambda}^t$, $\mathbf{h}_i$, for $i=1,\ldots,n(N-k)$. The computation of the $i$th component is as follows
\begin{IEEEeqnarray*}{rCl}
  \mathbf{x}\cdot \mathbf{h}_i^t &=& \sum_{k=1}^n\sum_{j=0}^{N-1}h_{jn+k,i}\sigma_{k}(x_{j+1}) \\
   &=& \sum_{k=1}^n\sigma_{k}\left(\sum_{j=0}^{N-1}h_{jn+k,i}x_{j+1}\right) \\
   &=& \sigma_s\left(\sum_{j=1}^N h_{j,r}^c x_j\right) \\
   &=& \sigma_s\left(\tilde{\mathbf{x}}\cdot \mathbf{h}_r^c \right),
\end{IEEEeqnarray*}
where $r=\left\lfloor\frac{i}{n}\right\rfloor+1$, $s=i-(r-1)n$ and $ \mathbf{h}_r^c=(h_{1,r}^c,\ldots,h_{N,r}^c)^t$ is the $r$th column of $\mathbf{H}^t$.
It should be noted that the two last equations in the above follows from the fact that $\mathbf{h}_i$ is of the form
$\mathbf{h}_i=\left(\mathbf{h}_i^1,\mathbf{h}_i^2,\ldots,\mathbf{h}_i^{N}\right)^t$, where
\begin{equation*}
 \mathbf{h}_i^j= \left(\overbrace{0,\cdots,0}^{(s-1)-\mathrm{times}},h_{j,r}^c,\overbrace{0,\cdots,0}^{(n-s)-\mathrm{times}}\right),\quad j=1,\ldots,N.
\end{equation*}
Thus
\begin{IEEEeqnarray*}{rCl}
  \mathbf{x}\mathbf{H}_{\Lambda}^t &=& \left( \sigma_1(\tilde{\mathbf{x}}\cdot \mathbf{h}_1^c),
  \ldots,\sigma_n(\tilde{\mathbf{x}}\cdot \mathbf{h}_1^c),\ldots, \right. \\
 &&  \> \left. \sigma_1(\tilde{\mathbf{x}}\cdot\mathbf{h}_{N-k}^c), \ldots,\sigma_n(\tilde{\mathbf{x}}\cdot \mathbf{h}_{N-k}^c) \right) \\
   &=& (\sigma(\tilde{\mathbf{x}}\cdot \mathbf{h}_1^c),\ldots,\sigma(\tilde{\mathbf{x}}\cdot \mathbf{h}_{N-k}^c)) \\
   &=&  \sigma^{N-k}\left(\tilde{\mathbf{x}}\cdot \mathbf{h}_1^c,\ldots,\tilde{\mathbf{x}}\cdot \mathbf{h}_{N-k}^c\right)\\
   &=& \sigma^{N-k}\left(\tilde{\mathbf{x}}\mathbf{H}^t\right).
\end{IEEEeqnarray*}
Thus, $\rho((\sigma^{N-k})^{-1}(\mathbf{x}\mathbf{H}_{\Lambda}^t))=\mathbf{0}$ implies $\rho\left(\tilde{\mathbf{x}}\mathbf{H}^t\right)=\mathbf{0}$ which indicates $\rho(\tilde{\mathbf{x}})\in C$, and so $\mathbf{x}\in \Lambda$.
\end{IEEEproof}

Theorem~\ref{theorem_parity} is also valid in the non-binary case,  where the conditions of Theorem~\ref{theorem6} are fulfilled. The authors of \cite{ConstA} proposed Construction~A based on number fields for non-binary linear codes. They have used cyclotomic number fields $\mathbb{Q}(\xi_{p^r})$ and their maximal totally real subfields $\mathbb{Q}(\xi_{p^r}+\xi_{p^r}^{-1})$, $r\geq 1$, as examples for their construction method. Using their method for the binary case $p=2$ does not provide diversity and gives us the well known Construction A \cite{2} that we describe in this section. In Section~\ref{new_construction}, we
 propose a new method for using Construction A over number fields in the binary case.
\section{Monogenic Number Fields}\label{monogenic_sec}
In this section, we provide the required algebraic tools for developing Construction A lattices over a wider family of number fields: the monogenic number fields.
\begin{Definition}
Let $K$ be a number field of degree $n$ and $O_K$ be its ring of integers. If $O_K$, as a $\mathbb{Z}$-module, has a basis of the form $\left\{1,\alpha,\ldots,\alpha^{n-1}\right\}$, for some $\alpha\in O_K$, then $\alpha$ is a \emph{power generator}, the basis is a \emph{power basis} and $K$ is a \emph{monogenic number field}.
\end{Definition}

It is a classical problem in algebraic number theory to identify if a number field $K$ is monogenic or not. The quadratic and cyclotomic number fields are monogenic, but in general this is not the case. Dedekind \cite[p. 64]{narkiewicz} was the first to notice this by giving an example of a cubic field generated by a root of $t^3- t^2-2t-8$. The existence of a power generator simplifies the arithmetic in $O_K$. For instance, if $K$ is monogenic, then the task of factoring
$pO_K$ into prime ideals over $O_K$, which is a difficult task in general, reduces to factoring the minimal polynomial of $\alpha$ over $\mathbb{F}_p$, which is significantly easier.

The proposed framework of \cite{ConstA} for developing Construction A lattices assumes that the number field $K$ is a Galois extension of $\mathbb{Q}$. Therefore, our construction method based on monogenic number fields is not a special case of their method since there exist examples of number fields which are monogenic without being Galois extensions. For example let $K = \mathbb{Q}(\alpha)$, where $\alpha^3= 2$ and $\alpha$ is the real cube root of $2$. Then it is proved that $O_K=\mathbb{Z}[\alpha]$ \cite[p. 67]{serglang} and $K$ is monogenic. However, it is known that $\mathbb{Q}(\sqrt[3]{2})$ is not a Galois extension.

We start by gathering the proved results about monogenic number fields and then we propose an algorithmic method to develop Construction A over monogenic number fields. We present the results about the number fields with degree less than $4$. More details about monogenic number fields
can be found in \cite{gaal}.
\begin{Theorem}{\cite[p. 76]{serglang}}\label{Th1}
Let $m$ be a non-zero square-free integer and let $K=\mathbb{Q}(\sqrt{m})$.
If $m \equiv 2$ or $3$ $(\bmod \,\,4)$, then $O_K=\mathbb{Z}[\sqrt{m}]$ is a basis for $O_K$ over $\mathbb{Z}$. If $m = 1$ $(\bmod \,\,4)$, then $O_K=\mathbb{Z}[\frac{1+\sqrt{m}}{2}]$.
\end{Theorem}

Theorem~\ref{Th1} shows that all quadratic fields are monogenic. In the cubic case, however, these studies begin to get more complicated. In fact there are an infinite number of cyclic cubic
fields which have a power basis and also an infinite number which do not, and similarly for quartic fields \cite{Robertson}.

Let $A$ be a Dedekind ring, $K$ its quotient field, $E$ a
finite separable extension of $K$ of degree $n$, and $B$ the integral closure of $A$ in $E$. Let $W = \left\{w_1,\ldots, w_n\right\}$ be any set of $n$ elements of $E$. The discriminant is
\begin{equation}\label{disc_general_def}
  D_{E/K}(W)=\left(\det[\sigma_i(w_j)]_{i,j=1}^n\right)^2,
\end{equation}
where $\sigma_i$'s are $n$ distinct embeddings of $E$ in a given algebraic closure of $K$. If $M$ is a free module of rank $n$ over $A$ (contained in $E$), then we can define the discriminant of $M$ by means of a basis of $M$ over $A$. This notion is well defined up to the square of a unit in $A$.
\begin{Proposition}{\cite[p. 65]{serglang}}\label{prop1}
Let $M_1\subset M_2$ be two free modules of rank $n$ over $A$, contained in $E$. Then $D_{E/K}(M_1)$ divides $D_{E/K}(M_2)$. If $D_{E/K}(M_1)=uD_{E/K}(M_2)$ for some unit $u$ of $A$, then $M_1 = M_2$.
\end{Proposition}

It is useful to recall the following well-known result.
\begin{lemma}\cite[p. 1-2]{gaal}\label{lemma1}
Let $K$ be a number field of degree $n$ and $\alpha_1,\ldots,\alpha_n\in O_K$ be linearly independent over $\mathbb{Q}$ and set $Z_K=\mathbb{Z}[\alpha_1, \ldots , \alpha_n]$. Then
\begin{equation*}
D_{K/\mathbb{Q}}(\alpha_1, \ldots , \alpha_n)=J^2\cdot d_K,
\end{equation*}
where
$$J=[O_K^{+}:Z_K^{+}],$$
$O_K^{+}$ and $Z_K^{+}$ are the additive groups of the corresponding modules and $d_K$ is the discriminant of the field $K$.
\end{lemma}

Let $\alpha\in O_K$ be a primitive element of $K$, that is $K = \mathbb{Q}(\alpha)$. The index of $\alpha$ is
defined by the module index
\begin{equation}\label{index}
  I(\alpha)=[O_K^{+}:\mathbb{Z}[\alpha]^{+}].
\end{equation}
Obviously, $\alpha$ generates a power integral basis in $K$ if and only if $I(\alpha)=1$. The
minimal index of the field $K$ is defined by
$$\mu(K)=\min_{\alpha} I(\alpha),$$
where the minimum is taken over all primitive integers. The field index of $K$ is
$$m(K)=\min_{\alpha}\gcd I(\alpha),$$
where the greatest common divisor is also taken over all primitive integers of $K$. Monogenic fields have both $\mu(K)= 1$ and $m(K) = 1$, but $m(K) = 1$ is not sufficient for being monogenic.

Let $\left\{1, \omega_2, \ldots , \omega_n\right\}$ be an integral basis of $K$. Let
$$L(\mathbf{x})=x_1+x_2\omega_2+\cdots +x_n\omega_n,$$
with conjugates $L^{(i)}(\mathbf{x})=x_1+x_2\omega_2^{(i)}+\cdots +x_n\omega_n^{(i)}$, where $\omega_j^{(i)}=\sigma_i(\omega_j)$, for $i,j=1,\ldots, n$.
The form $L(\mathbf{x})=L(x_1,\ldots,x_n)$ is  the \emph{fundamental form} and
\begin{equation*}
  D_{K/\mathbb{Q}}\left(L(\mathbf{x})\right)=\prod_{1\leq i<j\leq n}\left(L^{(i)}(\mathbf{x})-L^{(j)}(\mathbf{x})\right)^2
\end{equation*}
is the \emph{fundamental discriminant}.
\begin{lemma}\cite[p. 2]{gaal}
We have
\begin{equation}\label{disc_form}
  D_{K/\mathbb{Q}}\left(L(\mathbf{x})\right)=\left(I(x_2,\ldots,x_n)\right)^2d_K,
\end{equation}
where $d_K$ is the discriminant of the field $K$ and $I(x_2,\ldots ,x_n)$ is a homogeneous form in $n - 1$ variables of degree $n(n- 1)/2$ with integer coefficients. This form $I(x_2,\ldots ,x_n)$ is  the index form corresponding to the integral basis $\left\{1, \omega_2, \ldots , \omega_n\right\}$.
\end{lemma}
\begin{lemma}
For any primitive integer of the form $\alpha=x_1+\omega_2x_2+\cdots +\omega_n x_n\in O_K$ we have
\begin{equation*}
  I(\alpha)=|I(x_2,\ldots ,x_n)|.
\end{equation*}
\end{lemma}
Indeed, the existence of a power basis is equivalent to the existence of a solution to $I(x_2,\ldots ,x_n) = \pm 1$.
\begin{Theorem}\cite[Theorem 7.1.8]{alaca}\label{disc_squre_free}
Let $K$ be an algebraic number field of degree $n$. Let $\alpha\in O_K$ be such that $K = \mathbb{Q}(\alpha)$. If $D_{K/\mathbb{Q}}(\alpha)$ is square-free, then $\left\{1,\alpha,\ldots,\alpha^{n-1}\right\}$ is an integral basis for $K$. Indeed, $K$ has a power integral basis.
\end{Theorem}

The computation of the discriminant for some families of polynomials with small degree is a straightforward job. Combining these computations along with the conditions of Theorem~\ref{disc_squre_free} gives some useful results.
\begin{Theorem}\cite[Theorems 7.1.10, 7.1.12, 7.1.15]{alaca}\label{deg3_deg4}
Let $a, b$ be integers such that
\begin{enumerate}
  \item $x^3 + ax + b$  is irreducible. Let $\theta\in \mathbb{C}$ be a root of $x^3 + ax + b$ so that $K = \mathbb{Q}(\theta)$ is a cubic field and $\theta\in O_K$. Then $D_{K/\mathbb{Q}}(\theta)=-4a^3 - 27b^2$. If  $D_{K/\mathbb{Q}}(\theta)$ is square-free or $D_{K/\mathbb{Q}}(\theta)=4m$, where $m$ is a square-free integer such that $m\equiv 2$ or $3$ $(\bmod\,\, 4)$, then $\left\{1, \theta, \theta^2\right\}$ is an integral basis for the cubic field $\mathbb{Q}(\theta)$.
  \item $x^4 + ax + b$  is irreducible. Let $\theta\in \mathbb{C}$ be a root of $x^4 + ax + b$ so that $K = \mathbb{Q}(\theta)$ is a quartic field and $\theta\in O_K$. Then $D_{K/\mathbb{Q}}(\theta)=- 27a^4+256b^3$. If  $D_{K/\mathbb{Q}}(\theta)$ is square-free, then $\left\{1, \theta, \theta^2,\theta^3\right\}$ is an integral basis for the quartic field $\mathbb{Q}(\theta)$.
\end{enumerate}
\end{Theorem}
\begin{Theorem}\cite[p. 176]{alaca}\label{Th3}
Let $K = \mathbb{Q}( \sqrt[3]{m})$, with $m \in \mathbb{Z}$ a cube-free number. Assume that $m = hk^2$ with $h, k > 0$ and $hk$ is square-free,  and let $\theta = m^{1/3}$. Then,
\begin{itemize}
  \item for $m^2 \not\equiv 1\pmod{9}$, we have $d_K = -27(hk)^2$, and the numbers $\left\{1,\theta,\theta^2/k\right\}$, form an integral basis of $\mathcal{O}_K$;
  \item for $m^2\equiv\pm 1 \pmod{9}$, we have $d_K = -3(hk)^2$, and the numbers
  $$\left\{1,\theta,\frac{k^2\pm k^2\theta+\theta^2}{3k}\right\},$$
  form an integral basis of $\mathcal{O}_K$.
\end{itemize}
\end{Theorem}
\noindent
This theorem shows that $\mathbb{Q}(\sqrt[3]{p})$ is monogenic for primes $p\equiv \pm 2,\pm5$ $(\bmod\,\, 9)$.

Let $a\in \mathbb{Z}$ be an arbitrary integer  and consider a root $\vartheta$ of the polynomial
\begin{equation}\label{cubic_simplest}
  f(x)=x^3 -ax^2 + (a + 3)x + 1.
\end{equation}
Then, $K = \mathbb{Q}(\vartheta)$ are  the \emph{simplest cubic fields} \cite{cubic}. This cubic equation has discriminant $D = (a^2 + 3a + 9)^2$ and if $a^2 + 3a + 9$ is prime, $D$ is also the discriminant of the field $\mathbb{Q}(\vartheta)$. Accordingly, we have $\mathcal{O}_K=\mathbb{Z}[\vartheta]$ \cite{cubic}.
More information about monogenic number fields with higher degrees can be found in \cite{gaal}.
\section{System Model and Performance Evaluation on Block-Fading Channels}\label{system_model}
In this section, we describe the system models that describe communication over fading and block-fading channels using algebraic lattices.
First, we describe communication over fading channels using algebraic lattices of the form $\sigma(O_K)$ where $K$ is a number field of degree $n$, $O_K$ the ring of integers of $K$ and $\sigma$ is the canonical embedding.  We also present the available design  criteria and performance measurements in fading channels.  Then, by using  Construction A lattices with an underlying $(N,k)$-linear code $\mathcal{C}$, this model is converted to a model that describes communication over a block-fading channel with fading block length $N$.

In communication over a flat fading channel, the received discrete-time  signal vector is given by
\begin{equation}\label{channel}
  \mathbf{y}_i=\mathbf{H_F}\mathbf{x}_i+\mathbf{z}_i,\quad i=1,\ldots, N,
\end{equation}
where $\mathbf{y}_i\in \mathbb{R}^n$ is the received $n$-dimensional real  signal vector, $\mathbf{x}\in\mathbb{R}^n$ is the transmitted $n$-dimensional real signal vector, $\mathbf{H_F} = \textrm{diag}(\mathbf{h})$ is  an $n\times n$ real matrix, $\mathbf{h} = (h_1,\ldots, h_n)\in \mathbb{R}^n$ is the flat fading diagonal matrix, and $\mathbf{z}_i \in \mathbb{R}^n$ is the noise vector whose samples are i.i.d. $\sim \mathcal{N}(0, \sigma^2)$. We define the signal-to-noise ratio (SNR) as $\rho = 1/\sigma^2$.

Let a frame be composed of $N$ modulation symbols, each one  with dimension $n$,  or composed of $nN$ channel uses. The case of complex signals obtained from $2$ orthogonal real signals can be similarly modeled by (\ref{channel}) by replacing $N$
with $N' = 2N$. In communication over a block-fading channel, we assume that the fading matrix $\mathbf{H_F}$ is constant during one frame and it changes independently from frame to frame. This corresponds to a block-fading channel with $n$ blocks \cite{blockfading}. We further assume perfect channel state information (CSI) at the receiver, i.e., the receiver perfectly knows the fading coefficients. Therefore, for a given fading realization, the channel transition probabilities are given by
\begin{equation}\label{channel_prob}
  p(\mathbf{y}|\mathbf{x},\mathbf{H_F})=(2\pi\sigma^2)^{-\frac{n}{2}}\exp\left(-\frac{1}{2\sigma^2}\|\mathbf{y}-\mathbf{H_F}\mathbf{x}\|^2\right).
\end{equation}
Moreover, we assume that the real fading coefficients follow a Nakagami$-m$ distribution
\begin{equation}\label{nakagami}
  p_h(x)=\frac{2m^mx^{2m-1}}{\Gamma(m)}e^{-mx^2},
\end{equation}
where $m > 0$ and $\Gamma(x)\triangleq \int_{0}^{+\infty}t^{x-1}e^{-t}dt $ is the Gamma function. In the literature $m \geq 0.5$ is usually considered  \cite{porakis}; however, the fading
distribution is well defined and reliable communication is possible for any
$0 < m < 0.5$. Define the coefficients $\gamma_i = h_i^2$  for $i =1,\ldots , n$, which correspond to the fading power gains with
probability density function  (PDF) $p_{\gamma}(x) = \frac{m^mx^{m-1}}{\Gamma(m)}e^{-mx}$ and cumulative distribution function (CDF) $P_{\gamma}(x) = 1 - \overline{\Gamma}(mx,m)$,
respectively, where $\overline{\Gamma}(a,x)\triangleq\frac{1}{\Gamma(a)}\int_{x}^{+\infty}t^{a-1}e^{-t}dt$
is the normalized incomplete Gamma function~\cite{statistic}.

Analyzing the Nakagami$-m$ fading channels, in which the fading coefficients have Nakagami$-m$ distribution,  recovers the analysis for other fading channels, including Rayleigh fading by setting $m = 1$ and Rician fading with parameter $\kappa$ by setting
$m = (\kappa + 1)^2/(2\kappa + 1)$~\cite{statistic2}.
\subsection{Multidimensional lattice constellations}
In communication using multidimensional constellations, the transmitted signal vectors $\mathbf{x}$ belong to an $n$-dimensional signal constellation $\mathcal{S}\subset \mathbb{R}^n$. We consider signal constellations $\mathcal{S}$ that are generated as a finite subset of points carved from the infinite lattice $\Lambda=\left\{\mathbf{uM}+\mathbf{x}_0|\mathbf{u}\in\mathbb{Z}^n\right\}$
with full rank generator matrix $\mathbf{M}\in \mathbb{R}^{n\times n}$ \cite{2}. For a given channel
realization, we define the faded lattice seen by the receiver as the lattice $\Lambda'$ whose generator matrix
is given by $ \mathbf{M}'= \mathbf{H_F}\mathbf{M}$. In order to simplify the labeling
operation, constellations are of the type $\mathcal{S} = \left\{\mathbf{Mu} + \mathbf{x}_0 |\mathbf{u}\in\mathbb{Z}_M^n
\right\}$, where $\mathbb{Z}_M = \left\{0, 1,\ldots,M - 1\right\}$ represents an
integer pulse-amplitude modulation (PAM) constellation, $\log_2(M)$ is the number of bits per dimension and $\mathbf{x}_0$ is an offset vector which minimizes the average transmitted energy. The rate of such constellations
is $R = \log_2(M)$ $\textrm{bit/s/Hz}$. This is usually referred to as full-rate uncoded transmission \cite{SLB}.
\subsection{Error performance of multidimensional lattice constellations over fading channels}\label{system_model_subsec_2}
The performance evaluation of multidimensional signal sets has attracted significant attention due to the signal
space diversity (SSD) that these constellations present \cite{SSD} and
the fact that they can be efficiently used to combat the signal
degradation caused by fading. The diversity order of a multidimensional signal set is the minimum number of distinct components
between any two constellation points. In other words, the diversity order is the minimum Hamming distance between
any two coordinate vectors of constellation points. To distinguish from other well-known types of diversity
(time, frequency, space, code) this type of diversity is called \emph{modulation diversity} or \emph{signal space diversity} (SSD) \cite{SSD}.

The design of such constellations has been extensively studied in \cite{alglattice1,SLB,viterbo,Pappi} and due to the difficulties in the analytical computation of the Voronoi cells of multidimensional constellations, their error performance has been evaluated only through approximations and bounds or only for specific lattice structures. The evaluation of multidimensional constellations can be done by considering the error performance of maximum likelihood (ML) decoder.  At a given $i$, $1\leq i\leq N$, a maximum likelihood decoder with perfect CSI makes an error whenever $\|\mathbf{y}_i-\mathbf{H_F}\mathbf{w}\|^2 \leq \|\mathbf{y}_i-\mathbf{H_F}\mathbf{x}_i\|$ for some $\mathbf{w}\in\mathcal{S}$, $\mathbf{w}\neq \mathbf{x}_i$. These inequalities define the so called decision region around $\mathbf{x}$. Under ML decoding, the frame error probability is then given by
\begin{IEEEeqnarray}{rCl}\label{prob_fading}
  P_f(\rho) = \mathbb{E}\left[P_f(\rho|\mathbf{h})\right]= \mathbb{E}\left[1-(1-P_s(\rho|\mathbf{h}))^N\right],
\end{IEEEeqnarray}
where $\mathbf{h}$ consists of the diagonal elements of $\mathbf{H_F}$, and $P_f(\rho|\mathbf{h})$ and $P_s(\rho|\mathbf{h})$ are the frame and $n$-dimensional symbol error probabilities for a given channel realization and SNR $\rho$, respectively. The average is also taken over the fading distribution.
For a given constellation $\mathcal{S}$, we have \cite{SLB}
\begin{equation*}
  P_s(\rho|\mathbf{h})=\mathbb{E}\left[P_s(\rho|\mathbf{x},\mathbf{h})\right]=\frac{1}{|\mathcal{S}|}\sum_{\mathbf{x}\in\mathcal{S}}\int_{\mathbf{y}\not\in\mathcal{V}(\mathbf{x},\mathbf{h})}p(\mathbf{y}|\mathbf{x},\mathbf{h})d\mathbf{y},
\end{equation*}
where $\mathcal{V}(\mathbf{x},\mathbf{h})$ is the decision region or Voronoi region for a given multidimensional lattice constellation point $\mathbf{x}$ and fading $\mathbf{H_F}$. Computing the Voronoi regions and the exact error
probability is in general a very hard problem. An sphere lower bound (SLB) on $P_f$ has been proposed in \cite{SLB}. The SLB dates back to Shannon's work \cite{Shannon} and it has been
thoroughly investigated in the literature. However, it is not
generally a reliable lower bound for the important practical cases of finite lattice constellations \cite{Pappi}. Therefore, another lower bound called multiple sphere lower bound (MSLB) is proposed in~\cite{Pappi} in which the concept of the sphere lower bound is extended to the case of finite signal sets.
\begin{Definition} The diversity order is defined as the asymptotic (for large SNR) slope of $P_f$ in a log-log scale, i.e.,
\begin{equation}\label{diversity}
  d\triangleq -\lim_{\rho \rightarrow \infty}\frac{\log P_f(\rho)}{\log \rho}.
\end{equation}
\end{Definition}
The diversity order is usually a function of the fading distribution and the signal constellation $\mathcal{S}$. It is proved that
the diversity order is the product of the signal space diversity and a parameter of the fading distribution \cite{SLB}.
\begin{Definition} A constellation $\mathcal{S}\subset \mathbb{R}^n$ has \emph{full diversity} if the ML decoder is able to decode correctly in presence of $n-1$ \emph{deep fades}\footnote{When the transmitter and receiver are surrounded by reflectors, a transmitted signal can traverse in multiple paths and the receiver sees the superposition of multiple copies of the transmitted signal with different attenuations, delays and phase shifts. This can result in either constructive or destructive interference, amplifying or attenuating the signal power of the receiver. Strong destructive interference is frequently referred to as a deep fade and may result in temporary failure of communication due to a severe drop in the channel signal-to-noise ratio.}.
\end{Definition}

In this paper, our focus is on infinite lattices and we recall the basics of the sphere lower bound for infinite
lattices $\mathcal{S}$ \cite{SLB,tarokh}. From the geometrical uniformity of lattices we have that $\mathcal{V}(\mathbf{x}, \mathbf{h}) = \mathcal{V}(\mathbf{w}, \mathbf{h})=\mathcal{V}_{\Lambda}(\mathbf{h})$, for all $\mathbf{x},\mathbf{w}\in\Lambda$. Therefore, we assume the transmission of the all-zero codeword, i.e., $\mathbf{x}_i =\mathbf{0}$, $i = 1,\ldots, N$. Then, the error probability is given by \cite{2}
\begin{equation}\label{P_f}
  P_{f}(\rho)=1-\mathbb{E}\left[\left(1-\int_{\mathbf{z}\not\in\mathcal{V}_{\Lambda}(\mathbf{h})}p(\mathbf{z})d\mathbf{z}\right)^N\right].
\end{equation}
Due to the circular symmetry of the Gaussian noise, replacing $\mathcal{V}_{\Lambda}(\mathbf{h})$ by an $n$-dimensional sphere $\mathcal{B}(\mathbf{h})$ of the same volume and radius $R(\mathbf{h})$ [6], yields the corresponding sphere lower bound on the
lattice performance \cite{SLB,tarokh}
\begin{equation}\label{p_SLB}
  P_f(\rho)\geq P_{SLB}(\rho)=1-\mathbb{E}\left[\left(1-\int_{\mathbf{z}\not\in\mathcal{B}(\mathbf{h})}p(\mathbf{z})d\mathbf{z}\right)^N\right].
\end{equation}
In \cite{SLB}, for normalization purposes, it is assumed that $\det(M)=1$ and sphere lower bound is obtained for normalized lattices. Here we present the sphere lower bound without this assumption. Equating the volume of $\mathcal{B}(\mathbf{h})$ which is \cite{2} $$\textrm{vol}(\mathcal{\mathcal{B}(\mathbf{h})})=\frac{\pi^{\frac{n}{2}}R(\mathbf{h})^n}{\Gamma\left(\frac{n}{2}+1\right)},$$
to the fundamental volume of the lattice given by $\textrm{vol}(\mathcal{V}_{\Lambda}(\mathbf{h})=\det(\mathbf{H_F}\mathbf{M})=\det(M)\prod_{i=1}^{n}h_i$ yields the sphere radius
\begin{equation}\label{R_h^2}
  R(\mathbf{h})^2=\frac{1}{\pi}\left(\Gamma\left(\frac{n}{2}+1\right)\det(M)\prod_{i=1}^n h_i\right)^{\frac{2}{n}}.
\end{equation}
The probability that the noise brings the received point outside the sphere $\mathcal{B}(\mathbf{h})$ is  expressed as \cite{SLB,Shannon,tarokh}
\begin{equation}\label{P_SLB_final}
  P_{SLB}(\rho)=1-\mathbb{E}\left[\left(1-\overline{\Gamma}\left(\frac{n}{2},\frac{R(\mathbf{h})^2}{2}\rho\right)\right)^N\right].
\end{equation}
\subsection{Optimal lattice constellations}
We need an estimate of the error probability of the above system to address the search for good constellations.
Consider the multidimensional constellation $\mathcal{S}\subset \Lambda$. Due to the geometrically uniformity of the lattice, we may simply write $P_e(\Lambda) = P_e(\Lambda |\mathbf{x})$ for any transmitted point $\mathbf{x}\in \Lambda$. Thus, $\mathbf{x}$ can be considered as the all zero vector.   By applying the union bound and taking into account the edge
effects of the finite constellation $\mathcal{S}$ compared to the infinite
lattice $\Lambda$,  we obtain an upper bound to the point error probability \cite{viterbo}
\begin{equation}\label{union bound}
  P_e(\mathcal{S})\leq P_e(\Lambda)\leq \sum_{\mathbf{x}\neq \mathbf{w}}P(\mathbf{x}\rightarrow \mathbf{w}),
\end{equation}
where $P(\mathbf{x}\rightarrow \mathbf{w})$ is the pairwise error probability, the probability that the received point $\mathbf{y}$ is closer to $\mathbf{w}$ than to $\mathbf{x}$ according to the  metric
\begin{equation}\label{metric}
  m(\mathbf{x}|\mathbf{y},\mathbf{h})=\sum_{i=1}^n|y_i-h_ix_i|^2,
\end{equation}
when $\mathbf{x}$ is transmitted. In \cite{viterbo}, using the Chernoff bounding technique, it is shown that
\begin{equation}\label{viterbo_bound}
  P(\mathbf{x}\rightarrow \mathbf{w})\leq \frac{1}{2}\prod_{x_i\neq w_i}\frac{4\sigma^2}{(x_i- w_i)^2}=\frac{(4\sigma^2)^{\ell}}{2d_p^{(\ell)}(\mathbf{x},\mathbf{w})^2},
\end{equation}
where $\ell=|\left\{1\leq i\leq n|x_i\neq w_i\right\}|$. Let us define $L=\min_{\mathbf{x}\neq \mathbf{w}\in \mathcal{S}}\left\{\ell\right\}$ as the diversity order.
Thus, the point error probability of a multidimensional signal set is essentially dominated by four factors and to improve
performance it is necessary to \cite{viterbo}
\begin{enumerate}
  \item minimize the average energy per constellation point;
  \item maximize the signal space diversity $L$;
  \item maximize the minimum $L$-product distance
  \begin{equation}\label{d_p_min}
    d_{p,\textrm{min}}^{(L)}=\prod_{x_i\neq y_i}^L |x_i-y_i|
  \end{equation}
  between any two points $\mathbf{x}$ and $\mathbf{y}$ in the constellation;
  \item minimize the product kissing number $\tau_p$ for the $L$-product distance, i.e., the total number of points at the
minimum $L$-product distance.
\end{enumerate}
To minimize the error probability, one should maximize the diversity
order $L$, i.e., have full diversity $L = n$. Algebraic lattices of the form $\sigma(O_K)$, where $O_K$ is the integers ring of a number field $K$, have diversity order $r_1+r_2$, where $(r_1,r_2)$ is the signature of $K$ \cite{alglattice1}.  Therefore, totally real algebraic lattices have full diversity. On the other hand, the rank of a lattice determines the number of vectors we get with a given power limit and smaller rank means  less constellation vectors.  Hence, it is preferable to look at full rank lattices. Next, we should decide which one of the full rank lattices has the biggest minimum product distance. For two lattices with the same minimum product distance, the one with smaller parallelotope has better performance. Due to  Theorem~\ref{vol_theorem}, in order to minimize the volume of algebraic lattices it suffices to minimize the discriminant.
\subsection{Poltyrev outage limit for lattices}
In the preceding subsections, we introduced the evaluation methods for finite multidimensional constellations, including  lattice constellations.   In order to evaluate infinite lattices over the AWGN channels \cite{QC_LDPC_lattice}, we usually employ Poltyrev limit \cite{polytrev}. Due to this limit, there exists a lattice $\Lambda$, with generator $\mathbf{G}_{\Lambda}$, of high enough dimension $n$ for which the transmission error probability over the AWGN channel decreases  to an arbitrary low value if and only if $\sigma^2<\sigma_{max}^2$, where $\sigma^2$ is the noise variance per dimension, and  $\sigma_{max}^2$ is the Poltyrev threshold which is given by
\begin{equation}\label{Poltyrev_AWGN}
  \sigma_{max}^2=\frac{\left|\det(\mathbf{G}_{\Lambda})\right|^{\frac{2}{n}}}{2\pi e}.
\end{equation}
Using Poltyrev threshold, a \emph{Poltyrev outage limit} for lattices over block-fading channels is proposed in \cite{outage}. It is proved that Poltyrev outage limit has diversity $L$ for a channel with $L$ independent block fadings, i.e., Poltyrev outage limit has full diversity \cite{outage}. Using our notations through this paper, for a fixed instantaneous fading $\mathbf{h}=(h_1,\ldots,h_n)$, Poltyrev threshold becomes \cite{outage}
\begin{equation}\label{Poltyrev_fading}
  \sigma_{max}^2(\mathbf{h})=\frac{\left|\det(\mathbf{G}_{\Lambda})\right|^{\frac{2}{nN}}\prod_{i=1}^n h_i^{\frac{2}{n}}}{2\pi e}.
\end{equation}
The decoding of the lattice with generator $\mathbf{G}_{\Lambda}$ is possible with a vanishing error probability if $\sigma^2<\sigma_{max}^2(\mathbf{h})$ \cite{polytrev,outage}.
Thus, for variable fading, an outage event occurs whenever
$\sigma^2>\sigma_{max}^2(\mathbf{h})$. The Poltyrev outage limit $P_{out}(\rho)$ is defined as follows \cite{outage}
\begin{IEEEeqnarray}{rCl}\label{p_out_poly}
  P_{out}(\rho)&=&\mathrm{Pr}\left(\sigma^2>\frac{\left|\det(\mathbf{G}_{\Lambda})\right|^{\frac{2}{nN}}\prod_{i=1}^n h_i^{\frac{2}{n}}}{2\pi e}\right)\nonumber\\
  &=&\mathrm{Pr}\left(\prod_{i=1}^n h_i^2<\frac{(2\pi e)^{n}}{\left|\det(\mathbf{G}_{\Lambda})\right|^{\frac{2}{N}}\rho^n}\right),
\end{IEEEeqnarray}
where $\left|\det(\mathbf{G}_{\Lambda})\right|=2^{nN+N-k}d_K^{\frac{N}{2}}$ for our lattices. The closed-form expression of $P_{out}(\rho)$ is not derived in \cite{outage}; however it can be  estimated numerically via Monte Carlo simulation. For a
given lattice, the frame error rate  after lattice decoding over a block-fading channel, can be compared to $P_{out}(\rho)$
to measure the gap in SNR and verify the diversity order.
\section{Construction A over Monogenic Number Fields}\label{new_construction}
In this section we give more precise information concerning the splitting of the primes over monogenic number fields that helps us to develop Construction A lattices over monogenic number fields.
\begin{Proposition}\cite[p. 27]{serglang}\label{main_theorem}
Let $A$ be a Dedekind ring with quotient field $K$. Let $E$ be a finite separable extension of $K$. Let $B$ be the integral closure of $A$ in $E$ and assume that $B = A[\alpha]$ for some element $\alpha$. Let $f$ be the irreducible
polynomial of $\alpha$ over $K$ and let $\mathfrak{p}$ be a prime of $A$. Consider $\overline{f}$ to be the reduction of $f\,\,(\bmod\,\,\mathfrak{p})$, and let
\begin{equation}\label{poly_decomposition}
 \overline{f}(x)=\overline{P_1}(x)^{e_1}\cdots \overline{P_r}(x)^{e_r},
\end{equation}
be the factorization of $\overline{f}$ into powers of irreducible factors over  $\overline{A}=A/\mathfrak{p}$. Then
\begin{equation}\label{prime_decomposition}
  \mathfrak{p}B=\mathfrak{P}_1^{e_1}\cdots \mathfrak{P}_r^{e_r},
\end{equation}
is the factorization of $\mathfrak{p}$ in $B$, so that $e_i$ is the ramification index of $\mathfrak{P}_i$ over $\mathfrak{p}$,
and we have
\begin{equation}\label{prime_formula}
  \mathfrak{P}_i=\mathfrak{p}B+P_i(\alpha)B,
\end{equation}
where $P_i\in A[x]$ is a polynomial with leading coefficient $1$ whose reduction $\bmod\,\, \mathfrak{p}$ is $\overline{P_i}$.
For each $i$, $\mathfrak{P}_i$ has residue class degree $[B/\mathfrak{P}_i:A/\mathfrak{p}]=d_i$, where $d_i=\textrm{deg}(\overline{P_i})$.
\end{Proposition}

In our case, $A=\mathbb{Z}$, $K=\mathbb{Q}$, $E=\mathbb{Q}(\alpha)$, $B=O_E=\mathbb{Z}[\alpha]$ and $\mathfrak{p}=2\mathbb{Z}$. Let $f$ be the minimal polynomial of $\alpha$ over $\mathbb{Q}$ and $\overline{f}=f\,\,(\bmod \,\, 2)$. Write the decomposition of $\overline{f}$ in $\mathbb{F}_2[x]$ as follows
\begin{equation*}
 \overline{f}(x)=\overline{P_1}(x)^{e_1}\cdots \overline{P_r}(x)^{e_r}.
\end{equation*}
Then, we have
\begin{equation*}
  2O_E=\mathfrak{P}_1^{e_1}\cdots \mathfrak{P}_r^{e_r},
\end{equation*}
where $\mathfrak{P}_j=2O_E+P_j(\alpha)O_E$, for $j=1,\ldots,n$. If there exists $\overline{P_i}$ such that $d_i=\textrm{deg}(\overline{P_i})=1$ then $O_E/\mathfrak{P}_i\simeq\mathbb{F}_2$. Now, we can define the map $\rho:O_E^N\rightarrow \mathbb{F}_2^N$ as componentwise reduction modulo $\mathfrak{P}_i$ and develop the Construction A lattice $\Gamma_{\mathcal{C}}=\rho^{-1}(\mathcal{C})$ for an $(N,k)$ linear code $\mathcal{C}$.

As the simplest case, we present our method for block-fading channels with two fading blocks, i.e.,  $n=2$.  We require  quadratic fields of the form $K=\mathbb{Q}(\sqrt{m})$, where $m$ is a positive square-free integer; these fields are totally real. Theorem~\ref{Th1} determines the structure of $\mathcal{O}_K$ for these number fields.
\begin{Theorem}\label{quad_theorem}
Let $K=\mathbb{Q}(\sqrt{m})$. Then, $2\mathcal{O}_K$ is totally ramified with $2\mathcal{O}_K\cong \mathfrak{P}^2$
when $m\equiv 2$ $(\bmod\,\,4)$ and $\mathfrak{P}=2\mathbb{Z}[\sqrt{m}]+\sqrt{m}\mathbb{Z}[\sqrt{m}]$, or
$m\equiv 3$ $(\bmod\,\,4)$, $\mathfrak{P}=2\mathbb{Z}[\sqrt{m}]+(\sqrt{m}+1)\mathbb{Z}[\sqrt{m}]$.
In both of these cases we have $\mathcal{O}_K/\mathfrak{P}\cong\mathbb{F}_2$.
If $m\equiv 1$ $(\bmod\,\,4)$, then $2\mathcal{O}_K$ is not totally ramified, but if $(m-1)/4$ is an even number, then  $2\mathcal{O}_K\cong \mathfrak{P}_1\mathfrak{P}_2$ and $\mathcal{O}_K/\mathfrak{P}_i\cong\mathbb{F}_2$, $i=1,2$, where $\mathfrak{P}_1=2\mathbb{Z}[\alpha]+\alpha\mathbb{Z}[\alpha]$ and $\mathfrak{P}_2=2\mathbb{Z}[\alpha]+(\alpha+1)\mathbb{Z}[\alpha]$, with $\alpha=(1+\sqrt{m})/2$.
\end{Theorem}
\begin{IEEEproof}
All quadratic fields of the form $\mathbb{Q}(\sqrt{m})$, where $m$ is a positive square-free integer, are monogenic and totally real. If $m\equiv 2$ or $3$ $(\bmod\,\,4)$, then $\alpha=\sqrt{m}$ is the generator of the power integral basis with minimal polynomial $f(x)=x^2-m$. In this case, $f$ always has a linear factor after reduction modulo $2$. Indeed, we have $\overline{f}(x)=x^2$ for even $m$'s and $\overline{f}(x)=(x+1)^2$ for odd $m$'s. If $m\equiv 1$ $(\bmod\,\,4)$ then $\alpha=(1+\sqrt{m})/2$ is the generator of power integral basis with minimal polynomial $f(x)=x^2-x-(m-1)/4$. It can be easily seen that in this case, $f$ has a linear factor after reduction modulo $2$  if and only if $(m-1)/4$ is an even number, i.e., $m\equiv 1$ $(\bmod\,\,8)$. In this case, $\overline{f}(x)=x(x+1)$. The rest of the proof follows from Proposition~\ref{main_theorem}.
\end{IEEEproof}

In all cases of Theorem~\ref{quad_theorem}, there is at least one prime ideal $\mathfrak{P}_i$ in $\mathcal{O}_K$ such that $\mathcal{O}_K/\mathfrak{P}_i\cong\mathbb{F}_2$.  Define the map $\rho:\mathcal{O}_K^N\rightarrow \mathbb{F}_2^N$ as componentwise reduction modulo $\mathfrak{P}_i$ and implement the Construction A lattice $\Gamma_{\mathcal{C}}=\rho^{-1}(\mathcal{C})$ for an $(N,k)$  binary LDPC code $\mathcal{C}$. Then, $\Lambda=\sigma^N(\Gamma_{\mathcal{C}})$ is a $1$-level LDPC lattice of diversity order $2$ in $\mathbb{R}^{2N}$.
\begin{Example}
We have seen that the simplest cubic fields $K=\mathbb{Q}(\vartheta)$ where $\vartheta$ is a root of the polynomial $f(x)=x^3 - ax^2 + (a + 3)x + 1$, is a totally real monogenic number field, when $a^2+3a+9$ is a prime number. Even though this condition holds, these families of number fields are useless for our case since for each $a\in\mathbb{Z}$, $x^3 -ax^2 + (a + 3)x + 1$ $(\bmod \,\,2)$ is one of the polynomials  $x^3+x^2+1$ or $x^3+x+1$ and both of these polynomials are irreducible over $\mathbb{F}_2$.

Another examples are $K=\mathbb{Q}(\theta)$ where $\theta$ has minimal polynomial of the form $x^3+ax+b$. In this case, if $-4a^3-27b^2$  or $(-4a^3-27b^2)/4$ are square free  then $K$ is monogenic. For example put $a=3$ and $b=2$. Then $-4a^3-27b^2=-4\cdot 59$ which is a square-free integer after dividing by $4$. Hence, $K=\mathbb{Q}(\theta)$ where $f(\theta)=\theta^3+3\theta+2=0$ is a monogenic number field. We have
$$\overline{f}(x)=x^3+x=x(x+1)^2.$$
Due to this factorization, each one of the primes $\mathfrak{P}_1=2\mathbb{Z}[\theta]+2\theta\mathbb{Z}[\theta]$ or $\mathfrak{P}_2=2\mathbb{Z}[\theta]+2(\theta+1)\mathbb{Z}[\theta]$  gives us $O_K/\mathfrak{P_i}\simeq \mathbb{F}_2$. It can be easily checked  that $\mathbb{Q}(\theta)$ is not totally real  which is the only problem about these family of cubic polynomials.

Pure cubic fields of the form $\mathbb{Q}(\sqrt[3]{p})$  are monogenic for primes $p\equiv \pm 2,\pm 5$ $(\bmod\,\,9)$. In this case the  factorization of $x^3-p$ always has a linear factor. Unfortunately,  all pure cubic
fields are complex.
$\hfill\square$\end{Example}

In the existing number fields of degree $3$, we did not find any parametric family for which both being totally real and having linear factor after reduction modulo $2$ hold. There are a lot of numerical studies for finding monogenic number fields. An excellent account is provided in the tables of~\cite[Section 11]{gaal} containing  all generators of power integral bases for
$130$ cubic fields with small discriminants (both positive and negative), cyclic quartic,  totally real and totally complex biquadratic number fields up to discriminants $10^6$ and $10^4$, respectively. Furthermore,  the five totally real cyclic sextic fields with smallest discriminants,  the $25$ sextic fields with an imaginary quadratic subfield with smallest absolute value of discriminants and their generators of power integral bases are also given in \cite{gaal}.

We could  generate  many examples of number fields with different degrees of which the aforementioned two conditions are fulfilled. We used SAGE \cite{sage} to generate these examples but most of these results were already included in~\cite{gaal}. Let us analyse the results of \cite{gaal} about totally real cubic fields.

The provided table in \cite[Tabel 11.1.1]{gaal} contains all power integral bases of totally real cubic fields of discriminants $49 \leq d_K \leq 3137$. The rows contain the following data: $d_K$, $(a_1,a_2,a_3)$, where $d_K$ is the discriminant of the field $K$, generated by a root $\vartheta$ of the polynomial $f(x)=x^3+a_1x^2+a_2x+a_3$, and $(I_0,I_1,I_2,I_3)$ coefficients of the index form equation. In most of these fields $\left\{l, \omega_2=\vartheta,\omega_3=\vartheta^2\right\}$ is an integral basis; if not, then an integral basis is given by $\left\{1, \omega_2, \omega_3\right\}$ with $\omega_2=(p_0+p_1\vartheta+p_2\vartheta^2)/p$, $\omega_3 = q_0+q_1\vartheta+q_2\vartheta^2)/q$ and the table includes the coefficients $\omega_2 = (p_0, p_1, p_2)/p$, $\omega_3 = (q_0, q_1, q_2)/q$.
Finally, the solutions $(x, y)$, of the index form equation are displayed. All generators of power integral bases of the field $K$ are of the form
$$\alpha=a\pm(x\omega_2+y\omega_3),$$
where $a\in \mathbb{Z}$ is arbitrary and $(x, y)$ is a solution of the index form equation. For  $\overline{a_i}\equiv a_i$ $(\bmod\,\, 2)$, $1\leq i\leq 3$, the polynomial $f$ admits a linear factor after reduction modulo $2$, in one of the following cases
\begin{enumerate}
  \item $\overline{a_3}=0$;
  \item $\overline{a_1}\neq 0$ and $\overline{a_2}=\overline{a_3}=0$;
  \item $\overline{a_1}\neq 0$, $\overline{a_2}\neq 0$ and $\overline{a_3}\neq 0$.
\end{enumerate}
Consequently, for the following values of discriminant in~\cite[Table 11.1.1]{gaal}, we obtain a full diversity Construction A lattice with binary linear codes as underlying code
$$\begin{array}{l}
  148,229,316,404,469,564,568,621,733,756, \\
  788,837,892,940,1016,1076,1101,1229,1300,1373, \\
  1384,1396,1436,1492,1524,1556,1573,1620,1708,1765, \\
  1901,1940,1944,1957,2021,2024,2101,2213,2296,2300, \\
  2349,2557,2597,2677,2700,2708,2804,2808,2836,2917, \\
  2981,3021,3028,
\end{array}$$
which is $53/93$ or $57\%$ of the cases.
\begin{Example}\label{example_div_3}
Consider the number field $K=\mathbb{Q}(\nu)$, where $\nu$ is the root of the polynomial $f(x)=ax^3+bx^2+cx+d=x^3-x^2-3x+1$. Due to the above discussion, $K$ is monogenic with $d_K=148$ and $\mathcal{O}_K=\mathbb{Z}[\nu]$. Since the discriminant of $f$, which is $\Delta=18abcd-4b^3d+b^2c^2-4ac^3-27a^2d^2=148$, is positive $f$ has $3$ real roots as follows
\begin{IEEEeqnarray*}{rCl}
  x_1 &=& \frac{-1}{3}\left(-1+\zeta^0C+\frac{\Delta_0}{\zeta^0C}\right)=-1.4812, \\
  x_2 &=& \frac{-1}{3}\left(-1+\zeta^1C+\frac{\Delta_0}{\zeta^1C}\right)=2.170086, \\
  x_3 &=& \frac{-1}{3}\left(-1+\zeta^2C+\frac{\Delta_0}{\zeta^2C}\right)=0.311107,
\end{IEEEeqnarray*}
in which $\Delta_0=b^2-3ac$, $\zeta=\frac{-1}{2}+\frac{\sqrt{3}}{2}i$ and
$$C=\sqrt[3]{\frac{\Delta_1\pm \sqrt{\Delta_1^2-4\Delta_0^3}}{2}},\quad\Delta_1=2b^3-9abc+27a^2d.$$
The integral basis of $K$ is generated by $\nu=x_1$ as $\left\{1,\nu,\nu^2\right\}$ and using the embeddings $\sigma_1$ that sends $x_1$ to $x_1$, $\sigma_2$ that sends $x_1$ to $x_3$ and $\sigma_3$ that sends $x_1$ to $x_2$, gives us
\begin{equation*}
  \mathbf{M}=\left[
               \begin{array}{ccc}
                 1 & 1 & 1 \\
                 x_1 & x_3 & x_2 \\
                 x_1^2 & x_3^2 & x_2^2 \\
               \end{array}
             \right],
\end{equation*}
as the generator matrix of the lattice $\sigma(\mathcal{O}_K)$. Decomposing  $\overline{f}(x)=f(x)\pmod{2}=x^3+x^2+x+1$ as $(x+1)^3$ admits the following decomposition
$$2\mathcal{O}_K=\mathfrak{P}^3,\quad \frac{\mathcal{O}_K}{\mathfrak{P}}\cong \mathbb{F}_2,$$
where $\mathfrak{P}=2\mathcal{O}_K+(x_1+1)\mathcal{O}_K$ is a prime ideal of $\mathcal{O}_K$.
It can be checked that $\left\{2,x_1+1,x_1^2-x_1-2\right\}$ is a $\mathbb{Z}$-basis for $\mathfrak{P}$. Thus, the generator matrix of the lattice $\sigma(\mathfrak{P})$ is
\begin{equation*}
  \mathbf{DM}=\left[
               \begin{array}{ccc}
                 2 & 2 & 2 \\
                 x_1+1 & x_3+1 & x_2+1 \\
                 x_1^2-x_1-2 & x_3^2-x_3-2 & x_2^2-x_2-2 \\
               \end{array}
             \right].
\end{equation*}
Now, we consider an $[N,k]$-LDPC code with parity check matrix $\mathbf{H}_{\mathcal{C}}$ and generator matrix $\mathbf{G}_{\mathcal{C}}=\left[
                            \begin{array}{cc}
                              \mathbf{I}_k & \mathbf{A} \\
                            \end{array}
                          \right]
$ that gives us the parity check  and generator matrices of the triple diversity $1$-level LDPC lattice $\Lambda=\sigma^N(\Gamma_{\mathcal{C}})$ as $\mathbf{M}_{\Lambda}$ and $\mathbf{H}_{\Lambda}$ in Theorem~\ref{theorem_parity}, respectively.
$\hfill\square$\end{Example}
\begin{Example}
Next, we analyze the totally real quartic number fields. First examples of such fields are simplest quartic fields which had power integral in only two cases; see \cite{gaal}. These two cases are $K_2=\mathbb{Q}(\vartheta_2)$ and $K_4=\mathbb{Q}(\vartheta_4)$ where $\vartheta_2$ is a root of $f(x)=x^4-2x^3-6x^2+2x+1$ and $\vartheta_4$ is a root of $f(x)=x^4-4x^3-6x^2+4x+1$. The integral bases and solutions of index form equations  with respect to these bases have been presented in \cite{gaal}. Let $\left\{1,\omega_1,\omega_2,\omega_3\right\}$ represent the integral bases of $K_2$ and $K_4$. The generators of the power integral basis of $K_2$ and $K_4$ are of the form $\alpha=a+x_1\omega_1+x_2\omega_2+x_3\omega_3$, where $a\in\mathbb{Z}$ is arbitrary and $(x_1,x_2,x_3)$ is a solution of the corresponding index form equations of $K_2$ and $K_4$. For each $\alpha$ of this form we need to find its minimal polynomial over $\mathbb{Q}$ to check whether its reduction modulo $2$ has linear factors or not.  The minimal polynomials have been computed using SAGE~\cite{sage} and are presented in \tablename~\ref{table1} and \tablename~\ref{table2} for $K_2$ and $K_4$, respectively.
\begin{table}[h]
\begin{center}
\small
\caption{Minimal polynomials of simplest quartic fields for $a=2$.}\label{table1}
\renewcommand{\arraystretch}{1.3}
\begin{tabular}{|c||l|}
  \hhline{-||-}
  $(x_1,x_2,x_3)$ & Minimal Polynomial \\
  \hhline{=::=}
  $(0,1,0)$ & $t^4-10t^3+25t^2-20t+5$  \\

  $(-1,1,0)$ & $t^4-8t^3+19t^2-12t+1$ \\

  $(6,5,-2)$ & $t^4-22t^3+169t^2-508t+421$ \\

  $(0,4,-1)$ & $t^4-20t^3+115t^2-260t+205$ \\

  $(-12,-4,3)$ & $t^4-4t^3-29t^2-44t-19$ \\

  $(-8,-3,2)$ & $t^4+6t^3+t^2-4t-1$ \\

  $(1,1,0)$& $t^4-12t^3+19t^2-8t+1$ \\

  $(-2,1,0)$ & $t^4-6t^3+t^2+4t+1$ \\

  $(-13,-9,4)$ & $t^4+36t^3+451t^2+2176t+2641$ \\

  $(4,2,-1)$ & $t^4-8t^3+19t^2-12t+1$  \\
  \hhline{-||-}
\end{tabular}
\end{center}
\end{table}
\begin{table}[h]
\begin{center}
\small
\caption{Minimal polynomials of simplest quartic fields for $a=4$.}\label{table2}
\renewcommand{\arraystretch}{1.3}
\begin{tabular}{|c||l|}
   \hhline{-||-}
  $(x_1,x_2,x_3)$ & Minimal Polynomial \\
  \hhline{=::=}
  $(3,2,-1)$ & $t^4-4t^3+2t^2+4t-1$  \\
  $(-2,-2,1)$ & $t^4-8t^2-8t-2$ \\
  $(4,8,-3)$ & $t^4-24t^3+208t^2-760t+958$ \\
  $(-6,-7,3)$ & $t^4+16t^3+88t^2+200t+158$ \\
  $(0,3,-1)$ & $t^4-8t^3+16t^2-8t-2$ \\
  $(1,3,-1)$ & $t^4-12t^3+50t^2-84t+47$ \\
  \hhline{-||-}
\end{tabular}
\end{center}
\end{table}
\begin{table*}[!t]
\centering
\small
\caption{Monogenic totally real bicyclic biquadratic number fields.}
\renewcommand{\arraystretch}{1.3}
\begin{tabular}{|c|c|c|c|c|c|c|}
 \hline
  $d_K$ & $m$ & $n$ & $l=(m,n)$ & $\alpha$ & Minimal Polynomial $f_\alpha$ & Linear\\
  &&&&&& factor in $\overline{f_{\alpha}}$\\
  \hhline{=======}
  $2304$ & $2$ & $3$ & $1$ & $\frac{\sqrt{2}+\sqrt{6}}{2}$ & $t^4-4t^2+1$ & Yes\\
  $7056$ & $7$ & $3$ & $1$ & $\frac{\sqrt{7}+\sqrt{3}}{2}$ & $t^4-5t^2+1$ &No\\
  $24336$ & $39$ & $3$ & $3$ & $-\sqrt{39}+2\frac{\sqrt{39}+\sqrt{3}}{2}+\frac{1+\sqrt{13}}{2}$ & $t^4-2t^3-11t^2+12t-3$ &No\\
  $57600$ & $6$ & $15$ & $3$ & $\frac{\sqrt{6}+\sqrt{10}}{2}$ & $t^4-8t^2+1$ & Yes\\
  $94846$ & $11$ & $7$ & $1$ & $\frac{\sqrt{11}+\sqrt{7}}{2}$ & $t^4-9t^2+1$ & No\\
  $313600$ & $10$ & $35$ & $5$ & $\frac{\sqrt{10}+\sqrt{14}}{2}$ & $t^4-12t^2+1$ &Yes\\
  $435600$ & $15$ & $11$ & $1$ & $\frac{\sqrt{11}+\sqrt{15}}{2}$ & $t^4-13t^2+1$ &No\\
  $659344$ & $203$ & $7$ & $7$ & $-\sqrt{203}+2\frac{\sqrt{203}+\sqrt{7}}{2}+\frac{1+\sqrt{203}}{2}$ & $t^4-2t^3-27t^2+28t-7$ &No\\
  \hline
\end{tabular}\\
\label{table3}
\end{table*}
We have that the minimal polynomials of the power generators of $K_2$ are equivalent to $t^4+t^2+1$ modulo $2$ which has no linear factor. For $K_4$, all of them are equivalent to either  $t^4$ or $t^4+1$ which have linear factors. It can be shown that  $d_{K_2}=2000$ and $d_{K_4}=2048$.

Totally real bicyclic biquadratic number fields are other examples. Using the algorithm described in \cite[Section 6.5.2]{gaal}, the minimal index $\mu(K)$ and all elements with minimal index in the $196$ totally real bicyclic biquadratic
number fields $K = \mathbb{Q}(\sqrt{m},\sqrt{n})$ with discriminant smaller than $ 10^6$ have been determined. The results are gathered in \cite[Table 11.2.5]{gaal}. In this table, the solutions of index form equation $I(x_2,x_3,x_4)=\mu(K)$ has been proposed. The cases with $\mu(K)=1$ are the cases that $K$ has power integral basis. In the cases that $K$ has a power integral basis with power generator $\alpha$,  we have computed the minimal polynomial and the results are summarized in \tablename~\ref{table3}.
$\hfill\square$\end{Example}

More quartic fields with certain signatures and Galois
groups are computed and gathered in \cite[Section 11.2.7]{gaal}.
The  tables in \cite[Section 11.2.7]{gaal} contain the following data. In the first column the discriminant of the field $K = \mathbb{Q}(\xi)$, the second column contains the coefficients $(a_1, a_2, a_3, a_4)$ of the minimal polynomial $f_{\xi}(x) = x^4 + a_1x^3 + a_2x^2 + a_3x + a_4$
of $\xi$. In the third column the minimal $m$ for which the index form equation
$I(x_2, x_3, x_4) = \pm m$ has solutions with $|x_2|, |x_3|, |x_4| < 10^{10}$. It is followed by an
integral basis of $K$ in case the integral basis is not the power basis.
Last column contains  the solutions $(x_2, x_3, x_4)$ with absolute values smaller than $ 10^{10}$ of the index
form equation $I(x_2,x_3,x_4)=\pm m$. We have collected the cases that $\mathbb{Q}(\xi)$ has a power integral basis and  $f_{\xi}$ admits a linear factor after reduction modulo $2$. We have presented these cases by their discriminants in the following lists:
\setdefaultleftmargin{0cm}{2cm}{}{}{}{}
\begin{enumerate}[1)]
\item totally real quartic fields with Galois group $A_4$
\begin{equation*}
\begin{array}{l}
  26569,33489,121801,165649,261121,270400,299209, \\
  346921,368449,373321,408321,423801,473344,\\
  502681,529984,582169,660969,877969;
\end{array}
\end{equation*}
\item totally real quartic fields with Galois group $S_4$
\begin{equation*}
\begin{array}{l}
  2777,6224,6809,7537,8468,10273,10889,11324, \\
  11344,11348,13676,13768,14656,15188,15529,15952.
\end{array}
\end{equation*}
\end{enumerate}
\section{Decoding of Full Diversity $1$-level  LDPC Lattices}\label{decod2}
In this section we propose a new decoder, which is based on sum-product algorithm of LDPC codes and sphere decoder \cite{sphere_dec} of low dimensional lattices, for full diversity $1$-level LDPC lattices. We also analyze the decoding complexity of the proposed algorithm.

Let $\mathcal{C}$ be an $(N,k)$-LDPC code and $O_K$ be the integers ring of a totally real number field $K$ of degree $n$. Let $\mathfrak{p}$ be a prime ideal of $O_K$ such that $O_K/\mathfrak{P}\simeq \mathbb{F}_2$. Also, consider $\sigma_1,\ldots,\sigma_n$ to be $n$ real embeddings of $K$. Every lattice vector $\mathbf{x}$ in $\sigma^N(\Gamma_{\mathcal{C}})=\sigma^N(\rho^{-1}(\mathcal{C}))$ has the following  form
\begin{IEEEeqnarray}{rCl}
  \mathbf{x} &=& \sigma^N(\mathbf{c}+\mathbf{p})\nonumber \\
   &=& \left(\sigma(c_1+p_1),\ldots,\sigma(c_N+p_N)\right) \nonumber\\
   &=& \left(\sigma_1(c_1+p_1),\ldots,\sigma_n(c_1+p_1),\ldots,\sigma_n(c_N+p_N)\right)\nonumber \\
   &=& \left(c_1+\sigma_1(p_1),\ldots,c_1+\sigma_n(p_1),\ldots,c_N+\sigma_n(p_N)\right)\nonumber\\
   &=& \mathbf{c}\otimes\underbrace{(1,\ldots,1)}_{n-times}+\sigma^N(\mathbf{p}), \label{x_general_form}
\end{IEEEeqnarray}
where $\otimes$ is the Kronecker product,  $\mathbf{c}\in \mathcal{C}$ and $\mathbf{p}\in\mathfrak{P}^N$. To simulate the operation of our decoding algorithm, we use Rayleigh block-fading channel model; see Section \ref{system_model}. Rayleigh fading is a reasonable model when there are many objects in the environment that scatter the radio signal before it arrives at the receiver. Due to the central limit theorem, if there is sufficiently much scatter, the channel impulse response is modelled as a Gaussian process. If the scatters have no dominant components, then such a process will have zero mean and phase evenly distributed between $0$ and $2\pi$ radians. Thus, the envelope of the channel response is Rayleigh distributed. Often, the gain and phase elements of such channel's distortion are  represented as complex numbers. In this case, Rayleigh fading is exhibited by a complex random variable with real and imaginary parts  modelled by independent and identically distributed zero-mean Gaussian processes.
With the aid of an in-phase/quadrature component interleaver \cite{ConstA,alglattice1}, it is possible to
remove the phase of the complex fading coefficients to obtain a real fading which is Rayleigh distributed and guarantee that the fading coefficients are independent from one real symbol to the next.

\begin{figure}[ht]
\centering
\definecolor{ccqqqq}{rgb}{0.8,0.,0.}
\definecolor{qqqqff}{rgb}{0.,0.,1.}
\begin{tikzpicture}[line cap=round,line join=round,>=triangle 45,x=0.55cm,y=0.55cm]
\clip(-2.5,-3.2) rectangle (13.3,3.2);
\draw [color=qqqqff,fill=qqqqff,fill opacity=1.0] (0.,2.) circle (0.1958295782955816cm);
\draw [color=qqqqff,fill=qqqqff,fill opacity=1.0] (2.,2.) circle (0.19582957829558148cm);
\draw [color=qqqqff,fill=qqqqff,fill opacity=1.0] (5.968123613353219,2.0036487133059837) circle (0.19582957829558065cm);
\draw [color=qqqqff,fill=qqqqff,fill opacity=1.0] (8.,2.) circle (0.19582957829558237cm);
\draw [color=qqqqff,fill=qqqqff,fill opacity=1.0] (10.,2.) circle (0.19582957829558148cm);
\draw [color=qqqqff,fill=qqqqff,fill opacity=1.0] (12.,2.) circle (0.19582957829558148cm);
\draw [color=qqqqff,fill=qqqqff,fill opacity=1.0] (-2.,2.) circle (0.1958295782955817cm);
\fill[color=ccqqqq,fill=ccqqqq,fill opacity=1.0] (-0.3104072285142831,-1.510438906256035) -- (-0.3104072285142831,-2.3104389062560333) -- (0.4895927714857198,-2.3104389062560333) -- (0.4895927714857198,-1.510438906256035) -- cycle;
\fill[color=ccqqqq,fill=ccqqqq,fill opacity=1.0] (1.7039177652434612,-1.503029812663545) -- (1.7039177652434612,-2.303029812663545) -- (2.503917765243461,-2.303029812663545) -- (2.503917765243461,-1.503029812663545) -- cycle;
\fill[color=ccqqqq,fill=ccqqqq,fill opacity=1.0] (3.6943731093482617,-1.5025459353625241) -- (3.6943731093482617,-2.3025459353625237) -- (4.494373109348263,-2.3025459353625237) -- (4.494373109348263,-1.5025459353625241) -- cycle;
\fill[color=ccqqqq,fill=ccqqqq,fill opacity=1.0] (7.513393340936015,-1.5104477980938014) -- (7.513393340936015,-2.3104477980938007) -- (8.313393340936013,-2.3104477980938007) -- (8.313393340936013,-1.5104477980938014) -- cycle;
\fill[color=ccqqqq,fill=ccqqqq,fill opacity=1.0] (9.614504018844382,-1.5113116207855286) -- (9.614504018844382,-2.3113116207855287) -- (10.414504018844383,-2.3113116207855287) -- (10.414504018844383,-1.5113116207855286) -- cycle;
\draw [color=ccqqqq] (-0.3104072285142831,-1.510438906256035)-- (-0.3104072285142831,-2.3104389062560333);
\draw [color=ccqqqq] (-0.3104072285142831,-2.3104389062560333)-- (0.4895927714857198,-2.3104389062560333);
\draw [color=ccqqqq] (0.4895927714857198,-2.3104389062560333)-- (0.4895927714857198,-1.510438906256035);
\draw [color=ccqqqq] (0.4895927714857198,-1.510438906256035)-- (-0.3104072285142831,-1.510438906256035);
\draw [color=ccqqqq] (1.7039177652434612,-1.503029812663545)-- (1.7039177652434612,-2.303029812663545);
\draw [color=ccqqqq] (1.7039177652434612,-2.303029812663545)-- (2.503917765243461,-2.303029812663545);
\draw [color=ccqqqq] (2.503917765243461,-2.303029812663545)-- (2.503917765243461,-1.503029812663545);
\draw [color=ccqqqq] (2.503917765243461,-1.503029812663545)-- (1.7039177652434612,-1.503029812663545);
\draw [color=ccqqqq] (3.6943731093482617,-1.5025459353625241)-- (3.6943731093482617,-2.3025459353625237);
\draw [color=ccqqqq] (3.6943731093482617,-2.3025459353625237)-- (4.494373109348263,-2.3025459353625237);
\draw [color=ccqqqq] (4.494373109348263,-2.3025459353625237)-- (4.494373109348263,-1.5025459353625241);
\draw [color=ccqqqq] (4.494373109348263,-1.5025459353625241)-- (3.6943731093482617,-1.5025459353625241);
\draw [color=ccqqqq] (7.513393340936015,-1.5104477980938014)-- (7.513393340936015,-2.3104477980938007);
\draw [color=ccqqqq] (7.513393340936015,-2.3104477980938007)-- (8.313393340936013,-2.3104477980938007);
\draw [color=ccqqqq] (8.313393340936013,-2.3104477980938007)-- (8.313393340936013,-1.5104477980938014);
\draw [color=ccqqqq] (8.313393340936013,-1.5104477980938014)-- (7.513393340936015,-1.5104477980938014);
\draw [color=ccqqqq] (9.614504018844382,-1.5113116207855286)-- (9.614504018844382,-2.3113116207855287);
\draw [color=ccqqqq] (9.614504018844382,-2.3113116207855287)-- (10.414504018844383,-2.3113116207855287);
\draw [color=ccqqqq] (10.414504018844383,-2.3113116207855287)-- (10.414504018844383,-1.5113116207855286);
\draw [color=ccqqqq] (10.414504018844383,-1.5113116207855286)-- (9.614504018844382,-1.5113116207855286);
\draw [color=qqqqff] (-2.,2.)-- (0.11278841319085048,-1.510438906256035);
\draw [color=qqqqff] (-2.,2.)-- (2.1167478825169566,-1.503029812663545);
\draw [color=qqqqff] (-2.,2.)-- (4.095386513805523,-1.5025459353625241);
\draw [color=qqqqff] (0.11278841319085048,-1.510438906256035)-- (0.,2.);
\draw (0.11278841319085048,-1.510438906256035)-- (2.,2.);
\draw [color=qqqqff] (2.1167478825169566,-1.503029812663545)-- (2.,2.);
\draw [color=qqqqff] (0.,2.)-- (7.899519721744715,-1.5104477980938014);
\draw [color=qqqqff] (2.,2.)-- (7.899519721744715,-1.5104477980938014);
\draw [color=qqqqff] (5.968123613353219,2.0036487133059837)-- (4.095386513805523,-1.5025459353625241);
\draw [color=qqqqff] (5.968123613353219,2.0036487133059837)-- (7.899519721744715,-1.5104477980938014);
\draw [color=qqqqff] (8.,2.)-- (10.011400871554992,-1.5113116207855286);
\draw [color=qqqqff] (10.,2.)-- (10.011400871554992,-1.5113116207855286);
\draw [color=qqqqff] (12.,2.)-- (10.011400871554992,-1.5113116207855286);
\draw [color=qqqqff] (8.,2.)-- (4.095386513805523,-1.5025459353625241);
\draw [color=qqqqff] (0.11278841319085048,-1.510438906256035)-- (0.7991076116607069,-0.9941380822780541);
\draw [color=qqqqff] (0.11278841319085048,-1.510438906256035)-- (1.,-1.);
\draw [color=qqqqff] (0.11278841319085048,-1.510438906256035)-- (1.1998102649971212,-1.0275299700560887);
\draw [color=qqqqff] (2.1167478825169566,-1.503029812663545)-- (2.490963259081123,-1.005268711537399);
\draw [color=qqqqff] (2.1167478825169566,-1.503029812663545)-- (2.290611932412916,-1.005268711537399);
\draw [color=qqqqff] (2.1167478825169566,-1.503029812663545)-- (2.858274024639503,-1.005268711537399);
\draw [color=qqqqff] (2.1167478825169566,-1.503029812663545)-- (2.6801839564899854,-1.005268711537399);
\draw [color=qqqqff] (4.095386513805523,-1.5025459353625241)-- (3.8544653433508667,-1.0108340261670716);
\draw [color=qqqqff] (4.095386513805523,-1.5025459353625241)-- (3.6318527581639697,-0.9997033969077269);
\draw [color=qqqqff] (4.095386513805523,-1.5025459353625241)-- (5.,-1.);
\draw (7.899519721744715,-1.5104477980938014)-- (7.400399641521534,-0.9949329469790268);
\draw [color=qqqqff] (7.899519721744715,-1.5104477980938014)-- (8.14816505113125,-1.0053185776680507);
\draw [color=qqqqff] (10.011400871554992,-1.5113116207855286)-- (9.290584426923871,-0.9949329469790268);
\draw [color=qqqqff] (10.011400871554992,-1.5113116207855286)-- (9.,-1.);
\draw [color=qqqqff] (10.,2.)-- (9.5,1.5);
\draw [color=qqqqff] (10.,2.)-- (9.208111902683846,1.5026964103991582);
\draw [color=qqqqff] (12.,2.)-- (11.301105197665144,1.5017383379988685);
\draw [color=qqqqff] (12.,2.)-- (11.112275548773802,1.5017383379988685);
\draw [color=qqqqff] (5.968123613353219,2.0036487133059837)-- (5.1382864465709135,1.5149480789360534);
\draw [color=qqqqff] (8.,2.)-- (7.899519721744715,-1.5104477980938014);
\draw (3.425402198207871,2.532672014252459) node[anchor=north west] {$\cdots$};
\draw (5.757036404563646,-1.3784563318927696) node[anchor=north west] {$\cdots$};
\draw [shift={(4.124599955571245,3.2766717586023306)},dash pattern=on 1pt off 1pt]  plot[domain=3.347099681359604:4.014876804903414,variable=\t]({1.*6.256246110487987*cos(\t r)+0.*6.256246110487987*sin(\t r)},{0.*6.256246110487987*cos(\t r)+1.*6.256246110487987*sin(\t r)});
\draw [shift={(4.42831078837296,5.337978284072746)},dash pattern=on 1pt off 1pt]  plot[domain=3.6205309447518905:4.386535873380058,variable=\t]({1.*7.243291973740491*cos(\t r)+0.*7.243291973740491*sin(\t r)},{0.*7.243291973740491*cos(\t r)+1.*7.243291973740491*sin(\t r)});
\draw [shift={(9.830998451556667,0.410112344773896)},dash pattern=on 1pt off 1pt]  plot[domain=2.9812589334493156:3.336702528985785,variable=\t]({1.*9.958728498696503*cos(\t r)+0.*9.958728498696503*sin(\t r)},{0.*9.958728498696503*cos(\t r)+1.*9.958728498696503*sin(\t r)});
\draw [shift={(11.134597424970197,0.5478873059648064)},dash pattern=on 1pt off 1pt]  plot[domain=2.9839433651101266:3.365903332315412,variable=\t]({1.*9.249297345877162*cos(\t r)+0.*9.249297345877162*sin(\t r)},{0.*9.249297345877162*cos(\t r)+1.*9.249297345877162*sin(\t r)});
\draw [shift={(9.312529092245304,14.53386859280516)},dash pattern=on 1pt off 1pt]  plot[domain=3.978163042885976:4.397857743331845,variable=\t]({1.*16.88405094651766*cos(\t r)+0.*16.88405094651766*sin(\t r)},{0.*16.88405094651766*cos(\t r)+1.*16.88405094651766*sin(\t r)});
\draw [shift={(11.48988739397793,17.276877721549788)},dash pattern=on 1pt off 1pt]  plot[domain=4.067539052318516:4.523559938129143,variable=\t]({1.*19.115452002123234*cos(\t r)+0.*19.115452002123234*sin(\t r)},{0.*19.115452002123234*cos(\t r)+1.*19.115452002123234*sin(\t r)});
\draw [shift={(10.602859478519747,9.802321202627384)},dash pattern=on 1pt off 1pt]  plot[domain=3.878231600333007:4.477824205365199,variable=\t]({1.*11.614017709480466*cos(\t r)+0.*11.614017709480466*sin(\t r)},{0.*11.614017709480466*cos(\t r)+1.*11.614017709480466*sin(\t r)});
\draw [color=qqqqff] (0.,2.)-- (0.5,1.5);
\draw [shift={(9.414802827072858,-4.224054951974129)},dash pattern=on 1pt off 1pt]  plot[domain=2.4432792197153597:2.8577461830518023,variable=\t]({1.*9.680814067502862*cos(\t r)+0.*9.680814067502862*sin(\t r)},{0.*9.680814067502862*cos(\t r)+1.*9.680814067502862*sin(\t r)});
\draw [shift={(11.982747083609802,-3.4010713159230543)},dash pattern=on 1pt off 1pt]  plot[domain=2.409553467973505:2.9053813851370984,variable=\t]({1.*8.086203935303072*cos(\t r)+0.*8.086203935303072*sin(\t r)},{0.*8.086203935303072*cos(\t r)+1.*8.086203935303072*sin(\t r)});
\draw [shift={(15.691129913738598,-10.472250617386686)},dash pattern=on 1pt off 1pt]  plot[domain=2.1233753735879874:2.483198970966547,variable=\t]({1.*14.653003610622202*cos(\t r)+0.*14.653003610622202*sin(\t r)},{0.*14.653003610622202*cos(\t r)+1.*14.653003610622202*sin(\t r)});
\draw [shift={(19.144522588106415,0.25736424307646744)},dash pattern=on 1pt off 1pt]  plot[domain=2.953284414293158:3.3328799987347333,variable=\t]({1.*9.309085494595958*cos(\t r)+0.*9.309085494595958*sin(\t r)},{0.*9.309085494595958*cos(\t r)+1.*9.309085494595958*sin(\t r)});
\draw [shift={(17.60594384281732,-0.04019476215796479)},dash pattern=on 1pt off 1pt]  plot[domain=2.9323136709165296:3.2919220554649073,variable=\t]({1.*9.820211391762237*cos(\t r)+0.*9.820211391762237*sin(\t r)},{0.*9.820211391762237*cos(\t r)+1.*9.820211391762237*sin(\t r)});
\draw [shift={(18.585169375661387,6.598172273142701)},dash pattern=on 1pt off 1pt]  plot[domain=3.4908190755674484:3.790722747189913,variable=\t]({1.*13.427564578510623*cos(\t r)+0.*13.427564578510623*sin(\t r)},{0.*13.427564578510623*cos(\t r)+1.*13.427564578510623*sin(\t r)});
\draw [shift={(17.580925338081716,5.204226720345692)},dash pattern=on 1pt off 1pt]  plot[domain=3.464337192078189:3.86728012680366,variable=\t]({1.*10.102534296367104*cos(\t r)+0.*10.102534296367104*sin(\t r)},{0.*10.102534296367104*cos(\t r)+1.*10.102534296367104*sin(\t r)});
\draw [shift={(19.806946995881276,-4.734232092754369)},dash pattern=on 1pt off 1pt]  plot[domain=2.4298319354548292:2.8237300410752284,variable=\t]({1.*10.310106850832467*cos(\t r)+0.*10.310106850832467*sin(\t r)},{0.*10.310106850832467*cos(\t r)+1.*10.310106850832467*sin(\t r)});
\draw (-2.315933643248822,3.3349547519232754) node[anchor=north west] {$\vartheta_1$};
\draw (-0.3352981346240238,3.3349547519232754) node[anchor=north west] {$\vartheta_2$};
\draw (1.670408709552987,3.309883416371062) node[anchor=north west] {$\vartheta_3$};
\draw (5.681822397907008,3.3349547519232754) node[anchor=north west] {$\vartheta_{N-3}$};
\draw (7.637386570979594,3.3349547519232754) node[anchor=north west] {$\vartheta_{N-2}$};
\draw (9.668164750708817,3.309883416371062) node[anchor=north west] {$\vartheta_{N-1}$};
\draw (11.623728923781401,3.309883416371062) node[anchor=north west] {$\vartheta_{N}$};
\draw (-0.2600841279673859,-2.2559530762202247) node[anchor=north west] {$\Phi_1$};
\draw (1.7706940517618375,-2.2559530762202247) node[anchor=north west] {$\Phi_2$};
\draw (3.7513295603866355,-2.2559530762202247) node[anchor=north west] {$\Phi_3$};
\draw (7.562172564322956,-2.2559530762202247) node[anchor=north west] {$\Phi_{k-1}$};
\draw (9.668164750708817,-2.2559530762202247) node[anchor=north west] {$\Phi_k$};
\end{tikzpicture}
\caption{Tanner graph for a full-diversity $1$-level LDPC lattice with regular $(3,6)$ LDPC code as underlying code.}\label{taner_graph}
\end{figure}
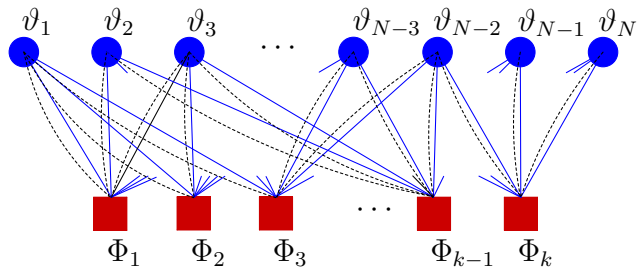
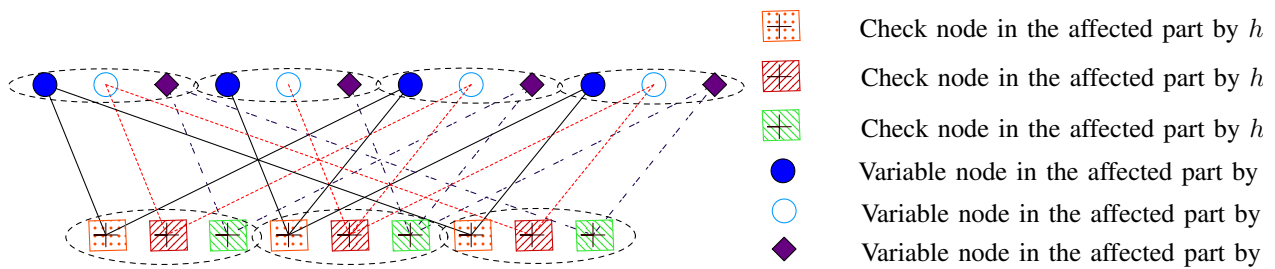
\begin{figure*}
  \centering
\definecolor{ffcctt}{rgb}{.2,0.9,0.2}
\definecolor{ccqqqq}{rgb}{0.8,0.,0.}
\definecolor{ffxfqq}{rgb}{1.,0.2980392156862745,0.}
\definecolor{xfqqff}{rgb}{0.1,0.,0.3}
\definecolor{ffqqqq}{rgb}{1.,0.,0.}
\definecolor{ttqqqq}{rgb}{0.2,0.,0.}
\definecolor{wwqqcc}{rgb}{0.4,0.,0.5}
\definecolor{qqzzff}{rgb}{0.,0.6,1.}
\definecolor{qqqqff}{rgb}{0.,0.,1.}
\begin{tikzpicture}[line cap=round,line join=round,>=triangle 45,x=0.8095238095238095cm,y=0.6666666666666666cm]
\clip(-3.,0.) rectangle (18.,6.);
\fill[color=ffxfqq,fill=ffxfqq,pattern=dots,pattern color=ffxfqq] (-1.28,1.28) -- (-1.26,0.68) -- (-0.66,0.7) -- (-0.68,1.3) -- cycle;
\fill[color=ffxfqq,fill=ffxfqq,pattern=dots,pattern color=ffxfqq] (1.7,1.28) -- (1.72,0.68) -- (2.32,0.7) -- (2.3,1.3) -- cycle;
\fill[color=ffxfqq,fill=ffxfqq,pattern=dots,pattern color=ffxfqq] (4.72,1.28) -- (4.74,0.68) -- (5.34,0.7) -- (5.32,1.3) -- cycle;
\fill[color=ccqqqq,fill=ccqqqq,pattern=north east lines,pattern color=ccqqqq] (-0.28,1.28) -- (-0.26,0.68) -- (0.34,0.7) -- (0.32,1.3) -- cycle;
\fill[color=ccqqqq,fill=ccqqqq,pattern=north east lines,pattern color=ccqqqq] (2.7,1.28) -- (2.72,0.68) -- (3.32,0.7) -- (3.3,1.3) -- cycle;
\fill[color=ccqqqq,fill=ccqqqq,pattern=north east lines,pattern color=ccqqqq] (5.72,1.28) -- (5.74,0.68) -- (6.34,0.7) -- (6.32,1.3) -- cycle;
\fill[color=ffcctt,fill=ffcctt,pattern=north west lines,pattern color=ffcctt] (0.7,1.28) -- (0.72,0.68) -- (1.32,0.7) -- (1.3,1.3) -- cycle;
\fill[color=ffcctt,fill=ffcctt,pattern=north west lines,pattern color=ffcctt] (3.7,1.28) -- (3.72,0.68) -- (4.32,0.7) -- (4.3,1.3) -- cycle;
\fill[color=ffcctt,fill=ffcctt,pattern=north west lines,pattern color=ffcctt] (6.72,1.28) -- (6.74,0.68) -- (7.34,0.7) -- (7.32,1.3) -- cycle;
\fill[color=ffxfqq,fill=ffxfqq,pattern=dots,pattern color=ffxfqq] (9.78,5.46) -- (9.8,4.86) -- (10.4,4.88) -- (10.38,5.48) -- cycle;
\fill[color=ccqqqq,fill=ccqqqq,pattern=north east lines,pattern color=ccqqqq] (9.78,4.46) -- (9.8,3.86) -- (10.4,3.88) -- (10.38,4.48) -- cycle;
\fill[color=ffcctt,fill=ffcctt,pattern=north west lines,pattern color=ffcctt] (9.78,3.48) -- (9.8,2.88) -- (10.4,2.9) -- (10.38,3.5) -- cycle;
\draw (-2.,4.)-- (-1.,1.);
\draw (4.,4.)-- (-1.,1.);
\draw [dash pattern=on 1pt off 1pt,color=ffqqqq] (-1.,4.)-- (0.,1.);
\draw [dash pattern=on 1pt off 1pt,color=ffqqqq] (5.,4.)-- (0.,1.);
\draw [dash pattern=on 1pt off 1pt on 2pt off 4pt,color=xfqqff] (0.,4.)-- (1.,1.);
\draw [dash pattern=on 1pt off 1pt on 2pt off 4pt,color=xfqqff] (6.,4.)-- (1.,1.);
\draw (1.,4.)-- (2.,1.);
\draw (4.,4.)-- (2.,1.);
\draw (7.,4.)-- (2.,1.);
\draw [dash pattern=on 1pt off 1pt,color=ffqqqq] (2.,4.)-- (3.,1.);
\draw [dash pattern=on 1pt off 1pt,color=ffqqqq] (5.,4.)-- (3.,1.);
\draw [dash pattern=on 1pt off 1pt,color=ffqqqq] (8.,4.)-- (3.,1.);
\draw [dash pattern=on 1pt off 1pt on 2pt off 4pt,color=xfqqff] (3.,4.)-- (4.,1.);
\draw [dash pattern=on 1pt off 1pt on 2pt off 4pt,color=xfqqff] (6.,4.)-- (4.,1.);
\draw [dash pattern=on 1pt off 1pt on 2pt off 4pt,color=xfqqff] (9.,4.)-- (4.,1.);
\draw (-2.,4.)-- (5.,1.);
\draw [dash pattern=on 1pt off 1pt,color=ffqqqq] (-1.,4.)-- (6.,1.);
\draw [dash pattern=on 1pt off 1pt on 2pt off 4pt,color=xfqqff] (0.,4.)-- (7.,1.);
\draw (5.,1.)-- (7.,4.);
\draw [dash pattern=on 1pt off 1pt,color=ffqqqq] (6.,1.)-- (8.,4.);
\draw [dash pattern=on 1pt off 1pt on 2pt off 4pt,color=xfqqff] (7.,1.)-- (9.,4.);
\draw [rotate around={0.:(-1.04,3.98)},dash pattern=on 2pt off 2pt] (-1.04,3.98) ellipse (1.2599998247177517cm and 0.22330847424983846cm);
\draw [rotate around={0.3819662047290258:(1.98,3.99)},dash pattern=on 2pt off 2pt] (1.98,3.99) ellipse (1.243392232342656cm and 0.22015964066186258cm);
\draw [rotate around={-0.7690246825780411:(4.97,3.98)},dash pattern=on 2pt off 2pt] (4.97,3.98) ellipse (1.2388085738182524cm and 0.2321843088289628cm);
\draw [rotate around={0.:(7.95,3.96)},dash pattern=on 2pt off 2pt] (7.95,3.96) ellipse (1.238436944394749cm and 0.23122060405169553cm);
\draw [color=ffxfqq] (-1.28,1.28)-- (-1.26,0.68);
\draw [color=ffxfqq] (-1.26,0.68)-- (-0.66,0.7);
\draw [color=ffxfqq] (-0.66,0.7)-- (-0.68,1.3);
\draw [color=ffxfqq] (-0.68,1.3)-- (-1.28,1.28);
\draw [color=ffxfqq] (1.7,1.28)-- (1.72,0.68);
\draw [color=ffxfqq] (1.72,0.68)-- (2.32,0.7);
\draw [color=ffxfqq] (2.32,0.7)-- (2.3,1.3);
\draw [color=ffxfqq] (2.3,1.3)-- (1.7,1.28);
\draw [color=ffxfqq] (4.72,1.28)-- (4.74,0.68);
\draw [color=ffxfqq] (4.74,0.68)-- (5.34,0.7);
\draw [color=ffxfqq] (5.34,0.7)-- (5.32,1.3);
\draw [color=ffxfqq] (5.32,1.3)-- (4.72,1.28);
\draw [color=ccqqqq] (-0.28,1.28)-- (-0.26,0.68);
\draw [color=ccqqqq] (-0.26,0.68)-- (0.34,0.7);
\draw [color=ccqqqq] (0.34,0.7)-- (0.32,1.3);
\draw [color=ccqqqq] (0.32,1.3)-- (-0.28,1.28);
\draw [color=ccqqqq] (2.7,1.28)-- (2.72,0.68);
\draw [color=ccqqqq] (2.72,0.68)-- (3.32,0.7);
\draw [color=ccqqqq] (3.32,0.7)-- (3.3,1.3);
\draw [color=ccqqqq] (3.3,1.3)-- (2.7,1.28);
\draw [color=ccqqqq] (5.72,1.28)-- (5.74,0.68);
\draw [color=ccqqqq] (5.74,0.68)-- (6.34,0.7);
\draw [color=ccqqqq] (6.34,0.7)-- (6.32,1.3);
\draw [color=ccqqqq] (6.32,1.3)-- (5.72,1.28);
\draw [color=ffcctt] (0.7,1.28)-- (0.72,0.68);
\draw [color=ffcctt] (0.72,0.68)-- (1.32,0.7);
\draw [color=ffcctt] (1.32,0.7)-- (1.3,1.3);
\draw [color=ffcctt] (1.3,1.3)-- (0.7,1.28);
\draw [color=ffcctt] (3.7,1.28)-- (3.72,0.68);
\draw [color=ffcctt] (3.72,0.68)-- (4.32,0.7);
\draw [color=ffcctt] (4.32,0.7)-- (4.3,1.3);
\draw [color=ffcctt] (4.3,1.3)-- (3.7,1.28);
\draw [color=ffcctt] (6.72,1.28)-- (6.74,0.68);
\draw [color=ffcctt] (6.74,0.68)-- (7.34,0.7);
\draw [color=ffcctt] (7.34,0.7)-- (7.32,1.3);
\draw [color=ffcctt] (7.32,1.3)-- (6.72,1.28);
\draw [color=ffxfqq] (9.78,5.46)-- (9.8,4.86);
\draw [color=ffxfqq] (9.8,4.86)-- (10.4,4.88);
\draw [color=ffxfqq] (10.4,4.88)-- (10.38,5.48);
\draw [color=ffxfqq] (10.38,5.48)-- (9.78,5.46);
\draw [color=ccqqqq] (9.78,4.46)-- (9.8,3.86);
\draw [color=ccqqqq] (9.8,3.86)-- (10.4,3.88);
\draw [color=ccqqqq] (10.4,3.88)-- (10.38,4.48);
\draw [color=ccqqqq] (10.38,4.48)-- (9.78,4.46);
\draw [color=ffcctt] (9.78,3.48)-- (9.8,2.88);
\draw [color=ffcctt] (9.8,2.88)-- (10.4,2.9);
\draw [color=ffcctt] (10.4,2.9)-- (10.38,3.5);
\draw [color=ffcctt] (10.38,3.5)-- (9.78,3.48);
\draw (11.18,2.68) node[anchor=north west] {\small Variable node in the affected part by $h_1$.};
\draw (11.22,1.84) node[anchor=north west] {\small Variable node in the affected part by $h_2$.};
\draw (11.22,1.06) node[anchor=north west] {\small Variable node in the affected part by $h_3$.};
\draw (11.2,5.54) node[anchor=north west] {\small Check node in the affected part by $h_1$.};
\draw (11.22,4.56) node[anchor=north west] {\small Check node in the affected part by $h_2$.};
\draw (11.22,3.54) node[anchor=north west] {\small Check node in the affected part by $h_3$.};
\draw [rotate around={0.:(3.01,0.96)},dash pattern=on 2pt off 2pt] (3.01,0.96) ellipse (1.3004908371026282cm and 0.365577705731889cm);
\draw [rotate around={0.:(-0.05,0.96)},dash pattern=on 2pt off 2pt] (-0.05,0.96) ellipse (1.300490837102624cm and 0.36557770573188786cm);
\draw [rotate around={0.:(6.09,0.98)},dash pattern=on 2pt off 2pt] (6.09,0.98) ellipse (1.3004908371026325cm and 0.3655777057318902cm);
\begin{scriptsize}
\draw [fill=qqqqff] (-2.,4.) circle (4.5pt);
\draw [color=qqzzff] (-1.,4.) circle (4.5pt);
\draw [fill=wwqqcc] (0.,4.) ++(-4.5pt,0 pt) -- ++(4.5pt,4.5pt)--++(4.5pt,-4.5pt)--++(-4.5pt,-4.5pt)--++(-4.5pt,4.5pt);
\draw [fill=qqqqff] (1.,4.) circle (4.5pt);
\draw [color=qqzzff] (2.,4.) circle (4.5pt);
\draw [fill=wwqqcc] (3.,4.) ++(-4.5pt,0 pt) -- ++(4.5pt,4.5pt)--++(4.5pt,-4.5pt)--++(-4.5pt,-4.5pt)--++(-4.5pt,4.5pt);
\draw [fill=qqqqff] (4.,4.) circle (4.5pt);
\draw [color=qqzzff] (5.,4.) circle (4.5pt);
\draw [fill=wwqqcc] (6.,4.) ++(-4.5pt,0 pt) -- ++(4.5pt,4.5pt)--++(4.5pt,-4.5pt)--++(-4.5pt,-4.5pt)--++(-4.5pt,4.5pt);
\draw [fill=qqqqff] (7.,4.) circle (4.5pt);
\draw [color=qqzzff] (8.,4.) circle (4.5pt);
\draw [fill=wwqqcc] (9.,4.) ++(-4.5pt,0 pt) -- ++(4.5pt,4.5pt)--++(4.5pt,-4.5pt)--++(-4.5pt,-4.5pt)--++(-4.5pt,4.5pt);
\draw [color=ttqqqq] (-1.,1.)-- ++(-4.5pt,0 pt) -- ++(9.0pt,0 pt) ++(-4.5pt,-4.5pt) -- ++(0 pt,9.0pt);
\draw [color=ttqqqq] (0.,1.)-- ++(-4.5pt,0 pt) -- ++(9.0pt,0 pt) ++(-4.5pt,-4.5pt) -- ++(0 pt,9.0pt);
\draw [color=ttqqqq] (1.,1.)-- ++(-4.5pt,0 pt) -- ++(9.0pt,0 pt) ++(-4.5pt,-4.5pt) -- ++(0 pt,9.0pt);
\draw [color=ttqqqq] (2.,1.)-- ++(-4.5pt,0 pt) -- ++(9.0pt,0 pt) ++(-4.5pt,-4.5pt) -- ++(0 pt,9.0pt);
\draw [color=ttqqqq] (3.,1.)-- ++(-4.5pt,0 pt) -- ++(9.0pt,0 pt) ++(-4.5pt,-4.5pt) -- ++(0 pt,9.0pt);
\draw [color=ttqqqq] (4.,1.)-- ++(-4.5pt,0 pt) -- ++(9.0pt,0 pt) ++(-4.5pt,-4.5pt) -- ++(0 pt,9.0pt);
\draw [color=ttqqqq] (5.,1.)-- ++(-4.5pt,0 pt) -- ++(9.0pt,0 pt) ++(-4.5pt,-4.5pt) -- ++(0 pt,9.0pt);
\draw [color=ttqqqq] (6.,1.)-- ++(-4.5pt,0 pt) -- ++(9.0pt,0 pt) ++(-4.5pt,-4.5pt) -- ++(0 pt,9.0pt);
\draw [color=ttqqqq] (7.,1.)-- ++(-4.5pt,0 pt) -- ++(9.0pt,0 pt) ++(-4.5pt,-4.5pt) -- ++(0 pt,9.0pt);
\draw [color=ttqqqq] (10.06,5.16)-- ++(-4.5pt,0 pt) -- ++(9.0pt,0 pt) ++(-4.5pt,-4.5pt) -- ++(0 pt,9.0pt);
\draw [color=ttqqqq] (10.06,4.16)-- ++(-4.5pt,0 pt) -- ++(9.0pt,0 pt) ++(-4.5pt,-4.5pt) -- ++(0 pt,9.0pt);
\draw [color=ttqqqq] (10.06,3.16)-- ++(-4.5pt,0 pt) -- ++(9.0pt,0 pt) ++(-4.5pt,-4.5pt) -- ++(0 pt,9.0pt);
\draw [fill=qqqqff] (10.12,2.28) circle (4.5pt);
\draw [color=qqzzff] (10.14,1.48) circle (4.5pt);
\draw [fill=wwqqcc] (10.14,0.72) ++(-4.5pt,0 pt) -- ++(4.5pt,4.5pt)--++(4.5pt,-4.5pt)--++(-4.5pt,-4.5pt)--++(-4.5pt,4.5pt);
\end{scriptsize}
\end{tikzpicture}
  \caption{Notation and diagram for the Tanner graph of a full diversity $1$-level  LDPC lattice for a block-fading channel with $3$ fading blocks.}\label{Ex_graph}
\hrulefill
\end{figure*}
Let $\mathbf{y}$ be the received  vector from Rayleigh block-fading channel with $n$ fading blocks and coherence time $N$
\begin{equation}\label{AWGN_output1}
\mathbf{y}^{t}=(\mathbf{I}_N\otimes\mathbf{H_F})\mathbf{x}^t+\mathbf{n}^t,
\end{equation}
where $\mathbf{H_F}=\textrm{diag}(|h_1|,\ldots,|h_n|)$ and the fading coefficients $h_i$ are complex Gaussian random
variables with variance $\sigma_b^2$, so that $|h_i|$ are Rayleigh distributed with parameter $\sigma_b^2$, for all $i = 1,\ldots , n$, and $\mathbf{n}=(\nu_1,\ldots,\nu_{nN})$, where $\nu_i\sim \mathcal{N}(0,\sigma^2)$ is the Gaussian noise, for $i=1,\ldots,nN$.
To simplify our decoding algorithm, we use the scaled and translated version of $\sigma^N(\Gamma_C)$ \cite[\S 20.5]{2}, \cite{sloane}. Hence, instead of $\mathbf{x}$, we use $\mathbf{x}'=2\mathbf{x}-(1,\ldots,1)$ as transmitted vector. Now, the received vector is
\begin{equation}\label{fading_output}
\mathbf{y}'^{t}=(\mathbf{I}_N\otimes\mathbf{H_F})\mathbf{x}'^{t}+\mathbf{n}^t=2(\mathbf{I}_N\otimes\mathbf{H_F})\mathbf{x}^{t}-(1,\ldots,1)^{t}+\mathbf{n}^{t}.
\end{equation}
First, we decode $\mathbf{p}$ and then we find $\mathbf{c}$. It is interesting to simulate iterative decoding of full-diversity $1$-level LDPC lattices for $n=2$, where the underlying code $\mathcal{C}$ is the $(3,6)$ ensemble (generalizations to other degree distributions and rates are treated similarly). The Tanner graph of this lattice is presented in \figurename~\ref{taner_graph}.
Transmitted information symbols are split into two classes: $N$ symbols are transmitted on $h_1$, while $N$ symbols are transmitted on $h_2$. Thus, there are two types of edges in \figurename~\ref{taner_graph}. Solid-line edges connect a variable node to a check node, both affected by $h_1$, and dashed-line edges connect a variable node to a check node, both affected by $h_2$.
Due to the structure of the parity check matrix of full-diversity $1$-level LDPC lattice in Theorem~\ref{theorem_parity}, there is no edge between the affected variable nodes by $h_1$ and the affected check nodes by $h_2$, conversely, there is no edge between the affected variable nodes by $h_2$ and the affected check nodes by $h_1$.
For each variable node $\vartheta_i$, $i=1,\ldots,N$, and check node $\Phi_j$, $j=1,\ldots,k$, we denote by $e_{i,j}$ and $e_{i,j}'$  the edges that connect $\vartheta_i$ to $\Phi_j$ in the affected part by $h_1$ and $h_2$, respectively. Indeed, $e_{i,j}$ is one of the solid-line edges while $e_{i,j}'$ is one of the dashed-line edges. Only one of these two edges with smaller fading effect, is chosen for decoding. This guarantees full-diversity under iterative message passing decoding \cite{rootLDPC}.
\begin{Example}
Let $\mathcal{C}$ be a binary LDPC code with parity check matrix $\mathbf{H}_{\mathcal{C}}$ as follows
\begin{equation}\label{H_C}
  \mathbf{H}_{\mathcal{C}}=\left[
                             \begin{array}{cccc}
                               1 & 0 & 1 & 0 \\
                               0 & 1 & 1 & 1\\
                               1 & 0 & 0 & 1\\
                             \end{array}
                           \right].
\end{equation}
A full diversity $1$-level LDPC lattice with diversity order $3$ has the following parity check matrix
 \begin{equation}\label{H_{land}}
   \mathbf{H}_{\Lambda}=\left[
                          \begin{array}{cccccccccccc}
                            1 & 0 & 0 & 0 & 0 & 0 & 1 & 0 & 0 & 0 & 0 & 0 \\
                            0 & 1 & 0 & 0 & 0 & 0 & 0 & 1 & 0 & 0 & 0 & 0 \\
                            0 & 0 & 1 & 0 & 0 & 0 & 0 & 0 & 1 & 0 & 0 & 0 \\
                            0 & 0 & 0 & 1 & 0 & 0 & 1 & 0 & 0 & 1 & 0 & 0 \\
                            0 & 0 & 0 & 0 & 1 & 0 & 0 & 1 & 0 & 0 & 1 & 0 \\
                            0 & 0 & 0 & 0 & 0 & 1 & 0 & 0 & 1 & 0 & 0 & 1 \\
                            1 & 0 & 0 & 0 & 0 & 0 & 0 & 0 & 0 & 1 & 0 & 0 \\
                            0 & 1 & 0 & 0 & 0 & 0 & 0 & 0 & 0 & 0 & 1 & 0 \\
                            0 & 0 & 1 & 0 & 0 & 0 & 0 & 0 & 0 & 0 & 0 & 1 \\
                          \end{array}
                        \right].
\end{equation}
The Tanner graph of this lattice is presented in \figurename~\ref{Ex_graph}. For decoding, we use the Tanner graph in \figurename~\ref{Ex_graph2} in which the solid line edges, corresponding to the edges with lower fading effect, are used in iterative decoding. Indeed, if we apply the Tanner graph of \figurename~\ref{Ex_graph} for our iterative decoding, the generated messages during the message passing iterations will not necessarily  preserve full diversity \cite{rootLDPC}.
\begin{figure}[ht]
  \centering
\definecolor{xfqqff}{rgb}{0.4980392156862745,0.,1.}
\definecolor{ffqqqq}{rgb}{1.,0.,0.}
\definecolor{ffxfqq}{rgb}{1.,0.4980392156862745,0.}
\definecolor{ttqqqq}{rgb}{0.2,0.,0.}
\definecolor{qqqqff}{rgb}{0.,0.,1.}
\begin{tikzpicture}[line cap=round,line join=round,>=triangle 45,x=0.8cm,y=0.7777777777777778cm]
\clip(-1.,-0.5) rectangle (9.,4.);
\fill[color=ffxfqq,fill=ffxfqq,pattern=dots,pattern color=ffxfqq] (0.16,0.64) -- (0.18,0.04) -- (0.78,0.06) -- (0.76,0.66) -- cycle;
\fill[color=ffxfqq,fill=ffxfqq,pattern=dots,pattern color=ffxfqq] (3.14,0.64) -- (3.16,0.04) -- (3.76,0.06) -- (3.74,0.66) -- cycle;
\fill[color=ffxfqq,fill=ffxfqq,pattern=dots,pattern color=ffxfqq] (6.16,0.64) -- (6.18,0.04) -- (6.78,0.06) -- (6.76,0.66) -- cycle;
\draw (-0.56,3.36)-- (0.44,0.36);
\draw (5.44,3.36)-- (0.44,0.36);
\draw (2.44,3.36)-- (3.44,0.36);
\draw (5.44,3.36)-- (3.44,0.36);
\draw (8.44,3.36)-- (3.44,0.36);
\draw (-0.56,3.36)-- (6.44,0.36);
\draw (6.44,0.36)-- (8.44,3.36);
\draw [color=ffxfqq] (0.16,0.64)-- (0.18,0.04);
\draw [color=ffxfqq] (0.18,0.04)-- (0.78,0.06);
\draw [color=ffxfqq] (0.78,0.06)-- (0.76,0.66);
\draw [color=ffxfqq] (0.76,0.66)-- (0.16,0.64);
\draw [color=ffxfqq] (3.14,0.64)-- (3.16,0.04);
\draw [color=ffxfqq] (3.16,0.04)-- (3.76,0.06);
\draw [color=ffxfqq] (3.76,0.06)-- (3.74,0.66);
\draw [color=ffxfqq] (3.74,0.66)-- (3.14,0.64);
\draw [color=ffxfqq] (6.16,0.64)-- (6.18,0.04);
\draw [color=ffxfqq] (6.18,0.04)-- (6.78,0.06);
\draw [color=ffxfqq] (6.78,0.06)-- (6.76,0.66);
\draw [color=ffxfqq] (6.76,0.66)-- (6.16,0.64);
\draw [shift={(6.6966666666666725,4.112222222222225)},dash pattern=on 1pt off 1pt,color=ffqqqq]  plot[domain=3.244883215773977:3.6818032001988934,variable=\t]({1.*7.295549971230143*cos(\t r)+0.*7.295549971230143*sin(\t r)},{0.*7.295549971230143*cos(\t r)+1.*7.295549971230143*sin(\t r)});
\draw [shift={(-5.6027272727272734,0.012424242424242404)},dash pattern=on 1pt off 1pt on 1pt off 4pt,color=xfqqff]  plot[domain=0.05745637380226927:0.5860447349910151,variable=\t]({1.*6.052715241923772*cos(\t r)+0.*6.052715241923772*sin(\t r)},{0.*6.052715241923772*cos(\t r)+1.*6.052715241923772*sin(\t r)});
\draw [shift={(8.890645161290331,3.8435483870967775)},dash pattern=on 1pt off 1pt,color=ffqqqq]  plot[domain=3.216413962157864:3.7102724538150067,variable=\t]({1.*6.468743466821224*cos(\t r)+0.*6.468743466821224*sin(\t r)},{0.*6.468743466821224*cos(\t r)+1.*6.468743466821224*sin(\t r)});
\draw [shift={(-2.7018749999999945,-0.02062499999999823)},dash pattern=on 1pt off 1pt on 1pt off 4pt,color=xfqqff]  plot[domain=0.06189296441496237:0.5816081443783221,variable=\t]({1.*6.153657766422335*cos(\t r)+0.*6.153657766422335*sin(\t r)},{0.*6.153657766422335*cos(\t r)+1.*6.153657766422335*sin(\t r)});
\draw [shift={(16.518461538461512,-20.770769230769186)},dash pattern=on 1pt off 1pt,color=ffqqqq]  plot[domain=2.0011928631657856:2.221238790965176,variable=\t]({1.*26.55233198285619*cos(\t r)+0.*26.55233198285619*sin(\t r)},{0.*26.55233198285619*cos(\t r)+1.*26.55233198285619*sin(\t r)});
\draw [shift={(-7.908085106382984,19.94014184397164)},dash pattern=on 1pt off 1pt on 1pt off 4pt,color=xfqqff]  plot[domain=5.115406494030344:5.390210467280204,variable=\t]({1.*21.28549927940293*cos(\t r)+0.*21.28549927940293*sin(\t r)},{0.*21.28549927940293*cos(\t r)+1.*21.28549927940293*sin(\t r)});
\draw [shift={(11.321142857142844,-2.727428571428564)},dash pattern=on 1pt off 1pt,color=ffqqqq]  plot[domain=2.338960579503949:2.768219520580502,variable=\t]({1.*8.46431497041356*cos(\t r)+0.*8.46431497041356*sin(\t r)},{0.*8.46431497041356*cos(\t r)+1.*8.46431497041356*sin(\t r)});
\draw [shift={(-2.2475,6.318333333333343)},dash pattern=on 1pt off 1pt on 1pt off 4pt,color=xfqqff]  plot[domain=5.474535520971676:5.915829886292362,variable=\t]({1.*8.237074259778858*cos(\t r)+0.*8.237074259778858*sin(\t r)},{0.*8.237074259778858*cos(\t r)+1.*8.237074259778858*sin(\t r)});
\draw [shift={(14.98457142857143,-13.214285714285715)},dash pattern=on 1pt off 1pt,color=ffqqqq]  plot[domain=1.9468663871309164:2.2755652670000455,variable=\t]({1.*17.819606115805396*cos(\t r)+0.*17.819606115805396*sin(\t r)},{0.*17.819606115805396*cos(\t r)+1.*17.819606115805396*sin(\t r)});
\draw [shift={(-4.279354838709662,18.892258064516103)},dash pattern=on 1pt off 1pt on 1pt off 4pt,color=xfqqff]  plot[domain=5.107068858505279:5.398548102805268,variable=\t]({1.*20.07568250634913*cos(\t r)+0.*20.07568250634913*sin(\t r)},{0.*20.07568250634913*cos(\t r)+1.*20.07568250634913*sin(\t r)});
\draw [shift={(14.29971428571428,-2.7131428571428504)},dash pattern=on 1pt off 1pt,color=ffqqqq]  plot[domain=2.338310607519766:2.7688694925646864,variable=\t]({1.*8.43915372969762*cos(\t r)+0.*8.43915372969762*sin(\t r)},{0.*8.43915372969762*cos(\t r)+1.*8.43915372969762*sin(\t r)});
\draw [shift={(0.7666666666666575,6.308888888888893)},dash pattern=on 1pt off 1pt on 1pt off 4pt,color=xfqqff]  plot[domain=5.474082215875044:5.916283191388994,variable=\t]({1.*8.220461673376803*cos(\t r)+0.*8.220461673376803*sin(\t r)},{0.*8.220461673376803*cos(\t r)+1.*8.220461673376803*sin(\t r)});
\draw [shift={(15.788571428571437,31.84)},dash pattern=on 1pt off 1pt,color=ffqqqq]  plot[domain=4.1912787833663145:4.423715604832898,variable=\t]({1.*32.83879089971346*cos(\t r)+0.*32.83879089971346*sin(\t r)},{0.*32.83879089971346*cos(\t r)+1.*32.83879089971346*sin(\t r)});
\draw [shift={(-9.046,-26.107333333333337)},dash pattern=on 1pt off 1pt on 1pt off 4pt,color=xfqqff]  plot[domain=1.0414057910866443:1.290403289932982,variable=\t]({1.*30.66489735475692*cos(\t r)+0.*30.66489735475692*sin(\t r)},{0.*30.66489735475692*cos(\t r)+1.*30.66489735475692*sin(\t r)});
\begin{scriptsize}
\draw [fill=qqqqff] (-0.56,3.36) circle (4.5pt);
\draw [fill=qqqqff] (2.44,3.36) circle (4.5pt);
\draw [fill=qqqqff] (5.44,3.36) circle (4.5pt);
\draw [fill=qqqqff] (8.44,3.36) circle (4.5pt);
\draw [color=ttqqqq] (0.44,0.36)-- ++(-4.5pt,0 pt) -- ++(9.0pt,0 pt) ++(-4.5pt,-4.5pt) -- ++(0 pt,9.0pt);
\draw [color=ttqqqq] (3.44,0.36)-- ++(-4.5pt,0 pt) -- ++(9.0pt,0 pt) ++(-4.5pt,-4.5pt) -- ++(0 pt,9.0pt);
\draw [color=ttqqqq] (6.44,0.36)-- ++(-4.5pt,0 pt) -- ++(9.0pt,0 pt) ++(-4.5pt,-4.5pt) -- ++(0 pt,9.0pt);
\end{scriptsize}
\end{tikzpicture}
  \caption{Tanner graph of a full diversity $1$-level LDPC lattice after choosing the edges with the least fading effect. }\label{Ex_graph2}
\end{figure}
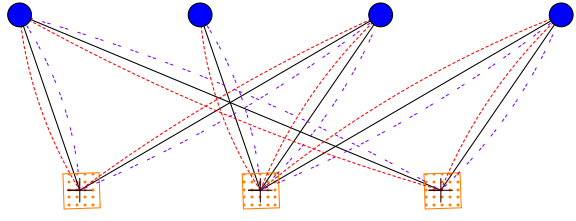
In \figurename~\ref{Ex_graph2}, the specified groups of nodes in  \figurename~\ref{Ex_graph} inside the dashed-line circles are merged.
$\hfill\square$\end{Example}

Define $\hat{\mathbf{p}}$, the estimation of $\mathbf{p}$, as follows
\begin{equation}\label{p_hat}
\hat{\mathbf{p}}=Q_{\Lambda_P'}\left(\mathbf{y}'^t\right),
\end{equation}
where $\Lambda_P'$ is the lattice with the following generator matrix $\mathbf{P}'$ and  $Q_{\Lambda_P'}(\mathbf{y}'^t)$ returns the $\textrm{argmin}_{\mathbf{z}\in\mathbb{Z}^{nN}}\|\mathbf{y}'^t-\mathbf{P}'\mathbf{z}^t\|^2$ with
$$\mathbf{P}'=2(\mathbf{I}_N\otimes\mathbf{H_F}\mathbf{P}^t),$$
in which $\mathbf{P}$ is the generator matrix of $\mathfrak{P}$ in $\mathbb{R}^n$. This decoding step seems to be a hard problem due to the high dimension of $\Lambda_P'$ which is $nN$. Here, we present a method which makes the complexity of this step affordable.   We use the following property of the Kronecker product in simplifying matrix equations. Consider three matrices $\mathbf{A}$, $\mathbf{B}$ and $\mathbf{X}$ such that $\mathbf{C}=\mathbf{AXB}$. Then \cite{horn1994topics}
\begin{equation}\label{vec_identity}
 ( \mathbf{B}^t\otimes \mathbf{A})\textrm{vec}(\mathbf{X})=\textrm{vec}(\mathbf{C}),
\end{equation}
where $\textrm{vec}(\mathbf{X})$ denotes the vectorization of the matrix $\mathbf{X}$ formed by stacking the columns of X into a single column vector. For each $\mathbf{z}=(z_{1},\ldots,z_{nN})\in\mathbb{Z}^{nN}$, we consider
\begin{eqnarray*}
  \mathbf{Z} &=& \left[
                   \begin{array}{cccc}
                     z_1 & z_{n+1} & \cdots & z_{(n-1)N+1} \\
                     z_2 & z_{n+2} & \cdots & z_{(n-1)N+2} \\
                     \vdots & \vdots & \ddots & \vdots \\
                     z_n & z_{2n} & \cdots & z_{nN} \\
                   \end{array}
                 \right].
\end{eqnarray*}
It is clear that $\textrm{vec}(\mathbf{Z})=\mathbf{z}^t$. By using (\ref{vec_identity}), we have
\begin{eqnarray*}
  \mathbf{P}'\mathbf{z}^t&=&2(\textrm{vec}\left(\mathbf{H_F}\mathbf{P}^t\mathbf{Z}\right))\\
  &=&\left(2\mathbf{z}_1^t\mathbf{P}\mathbf{H_F},\ldots,2\mathbf{z}_N^t\mathbf{P}\mathbf{H_F}\right)^t,
\end{eqnarray*}
where $\mathbf{z}_i$ is the $i$th column of $\mathbf{Z}$, for $i=1,\ldots,N$.
In a similar manner we can write
 \begin{eqnarray*}
  (\mathbf{I}_N\otimes\mathbf{H_F})\mathbf{x}^t&=&\left(\mathbf{x}_1^t\mathbf{H_F},\ldots,\mathbf{x}_N^t\mathbf{H_F}\right)^t,
\end{eqnarray*}
where $\mathbf{x}_i^t=\mathbf{x}((i-1)n+1:in)$, for $i=1,\ldots,N$. Consequently, we have
\begin{IEEEeqnarray*}{rCl}
  \|\mathbf{y}'^t-\mathbf{P}'\mathbf{z}^t\|^2 &=& \sum_{i=1}^N\| \mathbf{y}_i'^t-2\mathbf{H_FP}\mathbf{z}_i\|^2,
\end{IEEEeqnarray*}
where
\begin{IEEEeqnarray*}{rCl}
  \mathbf{y}_i'&=&2\mathbf{x}_i^t\mathbf{H_F}-(1,\ldots,1)+\mathbf{n}_i\\
  &=&2\mathbf{y}\left((i-1)n+1:in\right)-(1,\ldots,1),
\end{IEEEeqnarray*}
and $\mathbf{z}_i^t=\mathbf{z}\left((i-1)n+1:in\right)$. Indeed, it is enough to find $\textrm{argmin}_{\mathbf{z}_i\in\mathbb{Z}^{n}}\| \mathbf{y}_i'^t-2\mathbf{H_FP}\mathbf{z}_i\|^2$, for $i=1,\ldots,N$, which are $N$ instances of maximum likelihood (ML) decoding in dimension $n$. Since $n$ is the number of fading blocks, $n$ is small in comparison to the dimension of lattice $\Lambda=\sigma^N(\Gamma_C)$.
For computing the ML solutions, less complex methods exist; one of the most prominent ones being sphere decoding which is based on searching for the closest lattice point
within a given hyper-sphere \cite{sphere_dec}. In small dimensions, typically less than $100$, sphere decoding
is feasible after computing the Gram matrix \cite{sphere_dec}.
The steps for estimating $\hat{\mathbf{p}}$ is presented in Algorithm~\ref{decoding_step1}. The inputs of this algorithm are $\mathbf{P},\mathbf{H_F},\mathbf{y}$ and $\mathbf{R}$,
where
$$\mathbf{R}=\left[
               \begin{array}{rrrrr}
                 1 & 0 & 0 & 0 & 0 \\
                -1 & 1 & 0 & \cdots & 0\\
                -1 & 0 & 1 & \cdots & 0\\
                \vdots &\vdots &\vdots &\ddots &\vdots\\
                -1 & 0 & 0 & \cdots & 1
               \end{array}
             \right],$$
is the $n\times n$ noise reduction matrix that mitigates the effect of the noise created by $\mathbf{c}\otimes(1,\ldots,1)$ in the estimation of $\mathbf{p}$.  We call the matrix $\mathbf{R}^{(1 \leftrightarrow j)}$ in Algorithm~\ref{decoding_step1}, for $j=1,\ldots,n$, the $(1 \leftrightarrow j)$-\emph{row-column permutation} (RCP) of $\mathbf{R}$, which  is obtained by changing the position of the rows $j$ and $1$ in $\mathbf{R}$ (denote the obtained matrix by $\mathbf{R}'$) followed by changing the position of the columns $j$ and $1$ in  $\mathbf{R}'$.   The role of matrix $\mathbf{R}$ and its RCPs are vital because without multiplying by them,  the ML decoding in the lattice generated by $\mathbf{P}$  encounters with a noise with variance $\sigma_b'^2+4p_i(1-p_i)$, where $p_i=\mbox{Pr}\left\{c_i=1\right\}$, for $i=1,\ldots,N$ and $\sigma_b'$ is the variance of the product distribution of $h_i$ and $c_i$. The lattice generated by $\mathbf{P}$ has small volume and ML decoding can not afford such a big noise. The simulation results show that using our matrix $\mathbf{R}$ significantly improves the performance. The matrix $\mathbf{H_F}^{-1}=(h_{i,j}')$ in Algorithm~\ref{decoding_step1} is an $n\times n$ matrix given as follows
\begin{equation}\label{H_F_inv}
  h_{i,j}'=\left\{\begin{array}{ll}
                    \frac{1}{h_{i,j}}, & \mathrm{if}\,\, i=j \,\,  \mathrm{and} \,\,  h_{i,j}\neq 0,\\
                    0, &  \mathrm{otherwise}.
                  \end{array}
  \right.
\end{equation}
\begin{algorithm}
 \begin{algorithmic}[1]
 \Procedure{MI-ML}{$\mathbf{P},\mathbf{R},\mathbf{y},\mathbf{H_F}=\mbox{diag}(|h_1|,\ldots,|h_n|)$}
 \State $\hat{\mathbf{y}},\hat{\mathbf{h}},\hat{\mathbf{p}}\gets \mathbf{0}_{1\times N}$
 \For{$i=1:N$}
 \State $\mathbf{y}'_i\gets \mathbf{y}(n(i-1)+1:ni)$
 \State  $i_0 \gets \underset{{1\leq i\leq n}}{\textrm{arg}\max}\left(|h_1|,\ldots,|h_n|\right)$
               \State $\mathbf{y}_i^{ ''t}\gets \mathbf{R}^{(1 \leftrightarrow i_0)}\mathbf{H_F}^{-1}\mathbf{y}_i^{'t}$
                \State $\hat{\mathbf{z}}_i^t\gets \underset{{\mathbf{z}_i\in\mathbb{Z}^{n}}}{\textrm{arg}\min}\|
                \mathbf{y}_i^{ ''t}-2\mathbf{R}^{(1 \leftrightarrow i_0)}\mathbf{P}\mathbf{z}_i\|^2$

 \State $\mathbf{f}=\left(f_1,\ldots,f_n\right)\gets \mathbf{y}'_i-2\hat{\mathbf{z}}_i\mathbf{H_FP}$
 \State $i_m \gets \underset{{1\leq i\leq n} }{\textrm{arg}\max}\left(|f_1|,\ldots,|f_n|\right)$
 \State $\hat{\mathbf{p}}_i\gets 2\hat{\mathbf{z}}_i\mathbf{P}$
 \State $\hat{\mathbf{y}}(i)\gets \mathbf{y}'_i(i_m)-\mathbf{h}(i_m)\hat{\mathbf{p}}_i(i_m)$

 \State $\hat{\mathbf{h}}(i)\gets \mathbf{H_F}(i_m,i_m)$
 \State $\hat{\mathbf{p}}(i)\gets \hat{\mathbf{p}}_i(i_m)$
 \EndFor
 \State \textbf{return} $\hat{\mathbf{y}},\hat{\mathbf{h}},\hat{\mathbf{p}}$.
 \EndProcedure
 \end{algorithmic}
 \caption{\small{First step of decoding for full diversity $1$-level LDPC lattices}}
 \label{decoding_step1}
\end{algorithm}
\normalsize
After finding $\hat{\mathbf{p}}$, the estimation of $\mathbf{p}$, we need to find $\mathbf{c}$. After choosing the appropriate edges and discarding the remaining edges, our proposed algorithm is similar to the sum-product algorithm for LDPC codes in message passing structure \cite{sara}.
The sum-product algorithm iteratively computes an approximation of the MAP (maximum a posteriori probability) value for each code bit. The inputs are the log likelihood ratios (LLR) for the a priori message probabilities from each channel.
In the sequel, we introduce our method to estimate the vector of log likelihood ratios $\boldsymbol{\gamma}=(\gamma_1,\ldots, \gamma_N)$  for $1$-level LDPC lattices in the presence of perfect CSI.
We define the  vector of log likelihood ratios as
$\boldsymbol{\gamma}=(2\hat{\mathbf{h}}\circ \hat{\mathbf{y}})/\sigma^2$,
where $\circ $ is the Hadamard product or entrywise product.
Then, we input $\boldsymbol{\gamma}$ to the sum-product decoder of LDPC codes that gives us $\hat{\mathbf{c}}$. We convert $\hat{\mathbf{c}}$ to $\pm 1$ notation and we denote the obtained vector by $\hat{\mathbf{c}}'$.
The final decoded lattice vector is
\begin{equation*}
\hat{\mathbf{x}}=\hat{\mathbf{c}}'\otimes\overbrace{(1,\ldots,1)}^{n}+\hat{\mathbf{p}}.
\end{equation*}
Decoding error happens when $\hat{\mathbf{c}} \neq \mathbf{c}$ or $\hat{\mathbf{p}} \neq \mathbf{p}$.
\subsection{Decoding analysis}\label{decoding_analysis}
In this section, we prove that a $1$-level LDPC lattice with diversity $n$ achieves diversity $n-1$ under the  decoder proposed in the previous section. We also employ the  notations introduced in the previous section.   In the first part of our decoding algorithm, we have $N$ instances of optimal decoding, for the lattice generated by $\mathbf{P}$, over an $n$-block-fading channel. First, we assume that the transmitted codeword $\mathbf{c}$ in (\ref{x_general_form}) is the all zero codeword. In this case, our decoding problem is $N$ instances of optimal decoding over an $n$-block-fading channel with an additive noise with variance $\sigma_b^2$. The lattice generated by $\mathbf{P}$ comes from an algebraic number field and it has diversity order $n$. Thus, at high SNRs, i.e., when $\sigma_b^2\rightarrow 0$, optimal decoding of this lattice admits diversity order $n$. Now, we consider the general case that $\mathbf{c}=(c_1,\ldots,c_N)$ is not the all zero codeword. In this case, the purpose of the instance $i$ of our optimal decoding, for $i=1,\ldots,N$, is to obtain  $\mathbf{p}_i=(\sigma_1(p_i),\ldots,\sigma_n(p_i))=(p_{i,1},\ldots,p_{i,n})\in\mathbf{P}$ from the received vector of the form
$$\mathbf{y}_i=(h_1(p_{i,1}+c_i)+e_{i,1},\ldots,h_n(p_{i,n}+c_i)+e_{i,n})$$
in which $e_{i,j}\sim\mathcal{N}(0,\sigma_b^2)$, for $j=1,\ldots,n$. We consider $e_{i,j}'=h_jc_i+e_{i,j}$ as the effective noise that is not necessarily small  in high SNRs and we reach to an error floor in the performance curve. Without loss of generality, assume $h_1>0$ is the maximum of $\left\{h_1,\ldots,h_n\right\}$. Using  Step $6$ of Algorithm~\ref{decoding_step1} gives
\begin{IEEEeqnarray*}{rCl}
  &&\mathbf{y}_i'\\
  &=&(p_{i,1}+c_i+\frac{e_{i,1}}{h_1},p_{i,2}+c_i+\frac{e_{i,2}}{h_2},\ldots,p_{i,n}+c_i+\frac{e_{i,n}}{h_n})\mathbf{R}^t\\
  &=&(p_{i,1}+c_i+\frac{e_{i,1}}{h_1},p_{i,2}-p_{i,1}+e_{i,2}^{\prime\prime},\ldots,p_{i,n}-p_{i,1}+e_{i,n}^{\prime\prime}),
\end{IEEEeqnarray*}
where $e_{i,j}^{\prime\prime}=\frac{e_{i,j}}{h_j}-\frac{e_{i,1}}{h_1}$, for $j=2,\ldots,n$, is the Gaussian noise with zero mean and variance $\sigma_e^2=\left(\frac{h_1^2h_j^2}{h_1^2+h_j^2}\right)\sigma_b^2$. If the maximum of $\left\{h_1,\ldots,h_n\right\}$ occurs at $h_j$, with $j\neq 1$, we use $\mathbf{R}^{(1\leftrightarrow j)}$ instead of $\mathbf{R}$.
Now, let us consider $n-2$ deep fades as $h_3=h_4=\cdots=h_{n}=0$. In this case, we have
\begin{equation}\label{y_i_pime}
  \mathbf{y}_i'=(p_{i,1}+c_i+\frac{e_{i,1}}{h_1},p_{i,2}-p_{i,1}+e_{i,2}^{\prime\prime},e_{i,3}^{\prime\prime},\ldots,e_{i,n}^{\prime\prime})^t,
\end{equation}
which is equivalent to $n-2$ deep fades over the lattice vector $(p_{i,1},p_{i,2}-p_{i,1},\ldots,p_{i,n}-p_{i,1})$ in the  generated lattice by $\mathbf{RP}$. It should be noted that $\mathbf{R}$ and its RCPs are unimodular matrices and consequently, multiplication by these matrices generates equivalent lattices to the generated lattice by $\mathbf{P}$. Due to the ability of $\mathbf{P}$ in affording $n-2$ deep fades at high SNRs, under optimal decoding, we are able to decode $(p_{i,1},p_{i,2}-p_{i,1},\ldots,p_{i,n}-p_{i,1})$  in the  generated lattice by $\mathbf{RP}$. After multiplying by $\mathbf{R}^{-1}$, $\hat{\mathbf{p}}_i=(\hat{p}_{i,1},\hat{p}_{i,2},\ldots,\hat{p}_{i,n})$ is recovered correctly. After $N$ instances, we obtain $\hat{\mathbf{w}}=\sigma^N(\hat{\mathbf{p}})=(\hat{\mathbf{p}}_1,\ldots,\hat{\mathbf{p}}_N)$ as the estimation of $\sigma^N(\mathbf{p})$  in (\ref{x_general_form}). We also conclude from Equation (\ref{y_i_pime}) that using underlying LDPC codes with low maximum Hamming weight, lowers $\mathrm{Pr}\{c_i=1\}$, for $i=1,\ldots,N$,  that gives faster convergence of the error performance curve to its asymptotic slope. During Step $8$ and Step $9$ of Algorithm~\ref{decoding_step1}, the edge with smallest fading effect (or higher fading gain) are chosen among the $n$ equivalent edges that connect $\vartheta_i$ to its adjacent check nodes. Indeed, considering the output of Step $8$ at instance $i$, which is a vector of length $n$ with $j$th component as
\begin{equation*}
 f_j= \left\{\begin{array}{ll}
           2h_j\left(p_{i,j}-\hat{p}_{i,j}+\frac{1}{2}\right)+e_{i,j}, & \textrm{if}\,\,c_i=1, \\
           2h_j\left(p_{i,j}-\hat{p}_{i,j}-\frac{1}{2}\right)+e_{i,j}, & \textrm{if}\,\,c_i=0,
         \end{array}
  \right.
\end{equation*}
indicates that choosing $f_j$'s with higher absolute values  increases the  reliability in the estimation of log likelihood ratios.  During this edge discarding process, the nodes that decline the diversity order are removed and the remaining edges are not affected by deep fades. Hence, in presence of $n-2$ deep fades, the iterative decoding of $\mathbf{c}$  can be accomplished successfully over the obtained Tanner graph at high SNRs.

In order to discus the decoding complexity of the proposed algorithm, let us consider the complexity of the used optimal decoder in dimension $n$ as $f(n)$, which is a cubic polynomial for sphere decoder in high SNRs. Since our decoding involves $N$ uses of an optimal decoder in dimension $n$,  the complexity of our decoding method is $O(N\cdot f(n))+O(N\cdot d\cdot t)$ in which $t$ is the maximum number of iterations in the iterative decoding and $d$ is the average column degree of $\mathbf{H}_{\mathcal{C}}$. This complexity is dominated by $O(N\cdot d\cdot t)$ as $N$ is much greater than $n$.
\section{Numerical Results}\label{Numerical_Results}
\begin{center}
\begin{figure}[ht]
\centering
\includegraphics[width=4.5in]{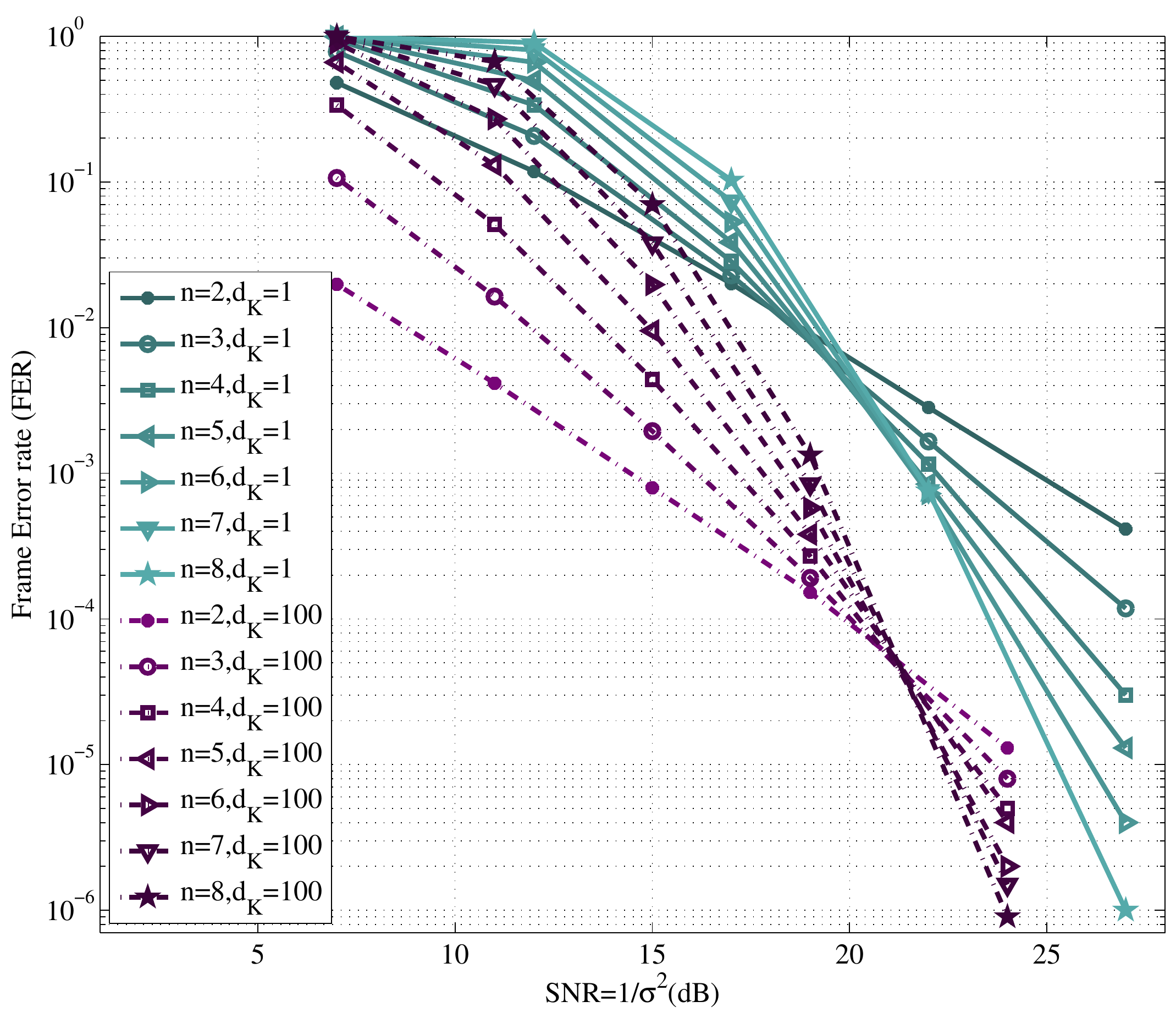}
\caption{\small{Poltyrev outage limit for $1$-level LDPC lattices with $[N,k]=[100,50]$ and different diversity orders.}}
\label{figsim0}
\vspace{-0.5cm}
\end{figure}
\end{center}

\begin{center}
\begin{figure}[ht]
\centering
\includegraphics[width=4.5in]{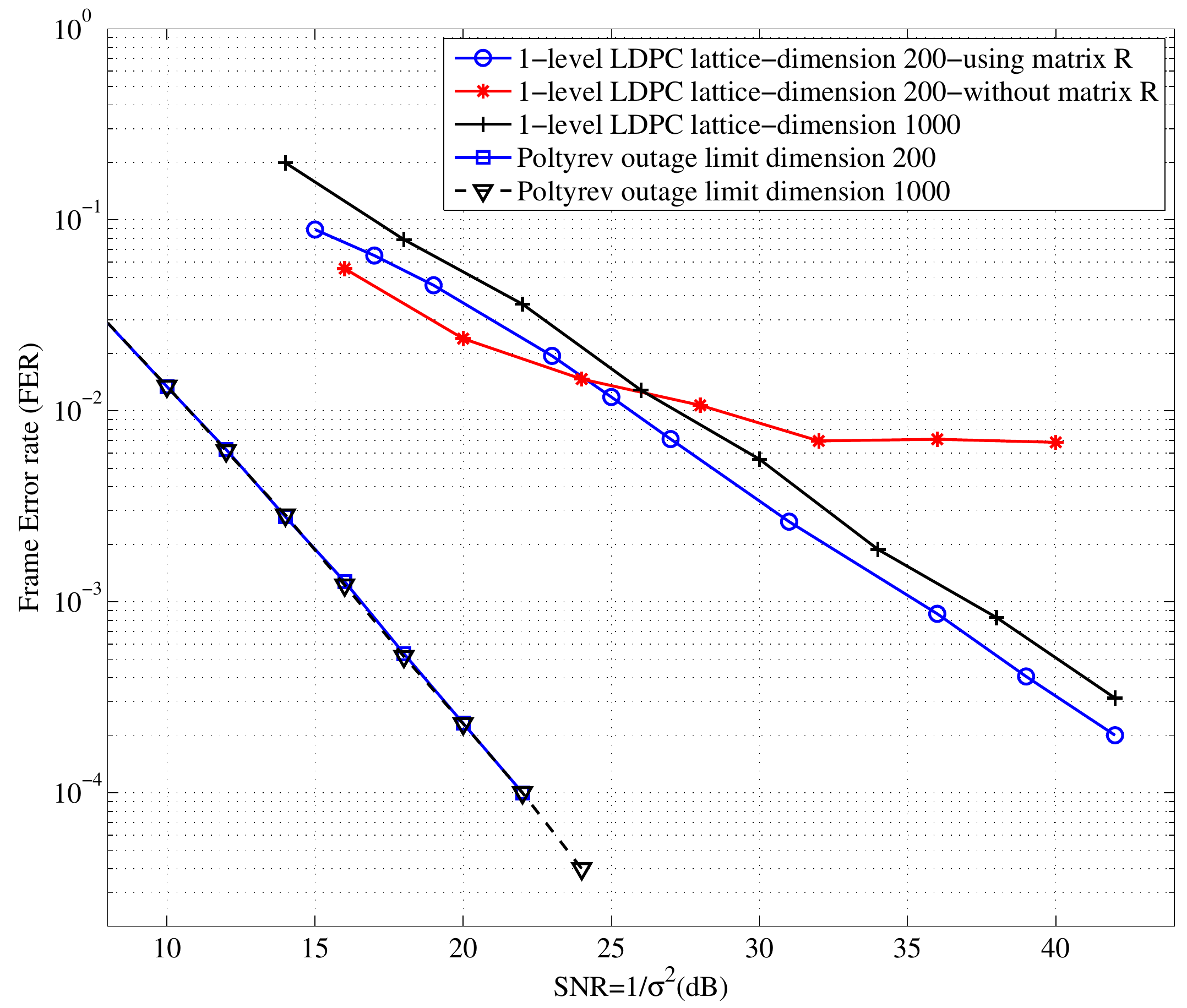}
\caption{\small{Decoding of double-diversity $1$-level LDPC lattices.}}
\label{figsim1}
\vspace{-0.5cm}
\end{figure}
\end{center}

\begin{center}
\begin{figure}[ht]
\centering
\includegraphics[width=4.5in]{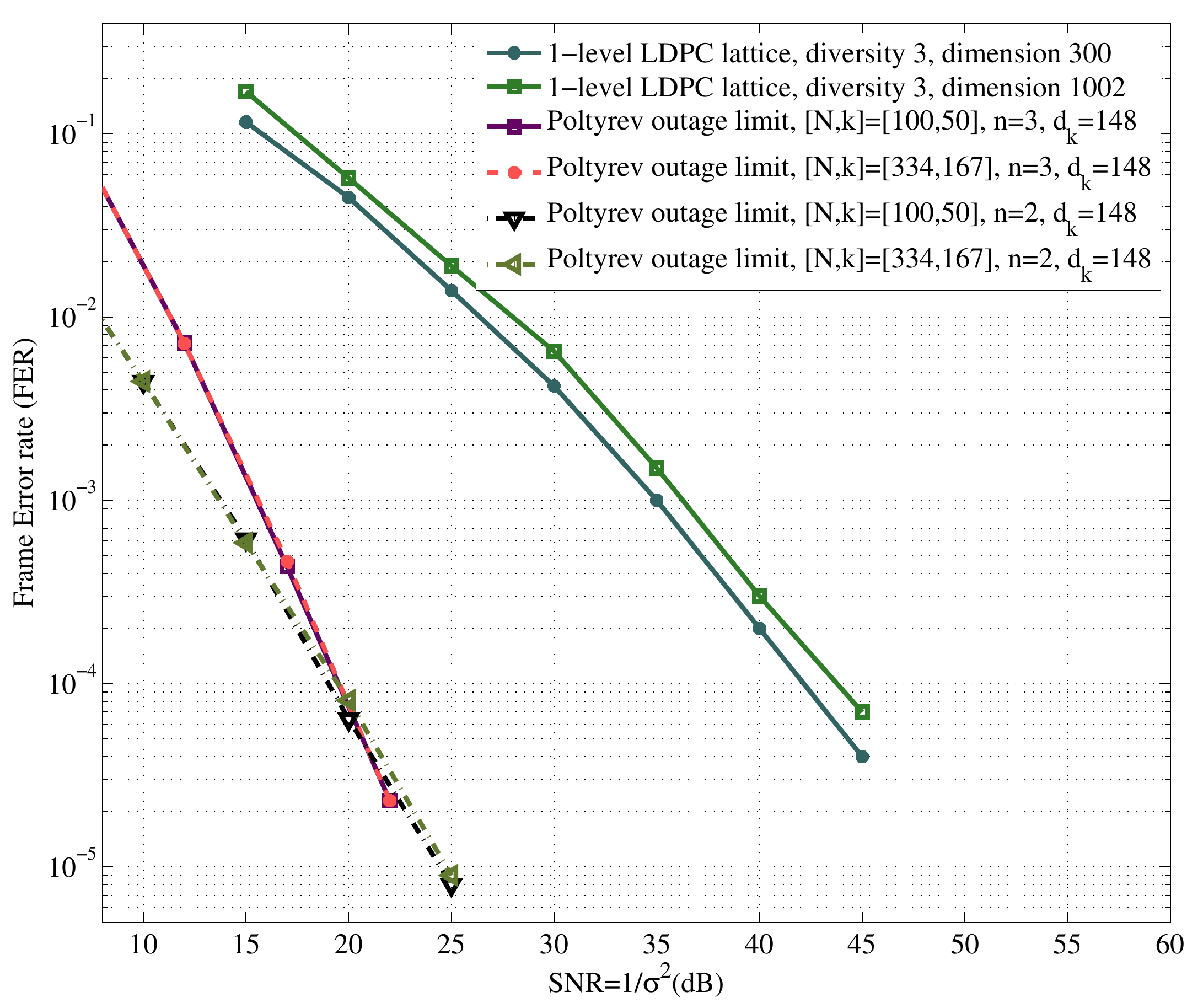}
\caption{\small{Decoding of triple-diversity $1$-level LDPC lattices.}}
\label{figsim2}
\vspace{-0.5cm}
\end{figure}
\end{center}

In this section, we present numerical results of simulating double diversity and triple diversity $1$-level LDPC lattices for block-fading channels.
Binary and randomly generated MacKay LDPC codes \cite{mackey} with parity-check matrices of size $50\times 100$, $250\times 500$ and $(167\times 334)$  are used in our simulations. Frame error rate (FER) performance of $1$-level LDPC lattices are plotted versus $\rho=1/\sigma^2$ in \figurename~\ref{figsim1}. In simulations we have used the construction of Theorem~\ref{main_theorem} with $m=10$ and the  decoding proposed in the previous section. The results for dimension $200$ are provided in presence and absence of multiplying by $\mathbf{R}$ in Algorithm~\ref{decoding_step1}.  We have compared the obtained results with the proposed Poltyrev outage limit in \cite{outage}. This outage limit is related to the fading distribution and determinant of the lattice which itself is related to $m$ and the rate of its underlying code. The Poltyrev outage limit of full diversity $1$-level LDPC lattices with different parameters and diversity orders are plotted in  \figurename~\ref{figsim0}. In \figurename~\ref{figsim2} we have presented the FER performance of triple diversity $1$-level LDPC lattices, obtained from Example~\ref{example_div_3} by employing $[100,50]$ and $[334,167]$ binary LDPC codes as underlying code. The Poltyrev outage limits with diversity order $2$ and $3$ are plotted for comparison. Due to the results of \figurename~\ref{figsim2}, triple diversity $1$-level LDPC lattices indicate diversity order $2$ under the proposed decoding algorithm in Section~\ref{decod2}, that confirms the proven result in Section~\ref{decoding_analysis}.
\vspace{-2cm}
\section{Conclusions}\label{conclusion}
In this paper, we propose full diversity $1$-level LDPC lattices on block-fading channels, based on algebraic number fields. The construction of $1$-level LDPC lattices with diversity order $2,3$ and $4$ is discussed through the paper. The framework for developing to higher orders of diversity is also provided.  In order to apply these structures in practical implementations, we propose a new low complexity decoding method for full diversity $1$-level LDPC lattices. The proposed decoder is based on optimal decoding in very small dimensions and iterative decoding. To implement the iterative part of our decoding algorithm, we propose the definition of a parity check matrix and Tanner graph for full diversity Construction A lattices.  The proposed decoding algorithm has complexity that grows linearly in the dimension of the lattice that makes it tractable to decode high-dimension $1$-level LDPC lattices on the block-fading channel. We also prove that the constructed LDPC lattices together with the proposed decoding method admit diversity order $n-1$ over an $n$-block fading channel.

\end{document}